\documentclass[a4paper]{spie}
\usepackage{graphicx}
\usepackage{subfig}
\usepackage{jnl_macros}
\usepackage{amsmath}
\usepackage{amsfonts}
\usepackage{amssymb}
\usepackage{booktabs}
\usepackage{multirow}

\newcommand{\imgdir}{.}

\title{Alternative approach to precision narrow-angle astrometry for
Antarctic long baseline interferometry}
\author{Yitping Kok$^*$\supit{a}, Michael J. Ireland\supit{b,c}, Aaron C.
  Rizzuto\supit{d}, Peter G. Tuthill\supit{e}, J. Gordon Robertson\supit{c,e},
  Benjamin A. Warrington\supit{f} and William J. Tango\supit{e}.
\skiplinehalf
\supit{a}MPE, Gie{\ss}enbachstra{\ss}e 1, 85748 Garching, Germany; \\
\supit{b}RSAA, Australian National University, Canberra, ACT 2611, Australia; \\
\supit{c}Australian Astronomical Observatory, PO Box 915, North Ryde, NSW 1670, Australia; \\
\supit{d}Department of Physics and Astronomy, Macquarie University, NSW 2109, Australia; \\
\supit{e}SIfA, School of Physics, University of Sydney, NSW 2006, Australia; \\
\supit{f}USRA-SOFIA, AFRC building 703, 2825 E. Ave. P, Palmdale, CA, 93550, USA. \\
}

\authorinfo{$^*$E-mail: yitping@mpe.mpg.de}

\begin{document}
\maketitle

\begin{abstract}
The conventional approach to high-precision narrow-angle astrometry using a long
baseline interferometer is to directly measure the fringe packet separation of a
target and a nearby reference star. This is done by means of a technique known
as phase-referencing which requires a network of dual beam combiners and laser
metrology systems. Using an alternative approach that does not rely on
phase-referencing, the narrow-angle astrometry of several closed binary stars
(with separation less than 2$''$), as described in this paper, was carried
out by observing the fringe packet crossing event of the binary systems. Such an
event occurs twice every sidereal day when
the line joining the two stars of the binary is
is perpendicular to the projected baseline of the interferometer. Observation of
these events is well suited for an interferometer in Antarctica. Proof of
concept observations were carried out at the Sydney University Stellar
Interferometer (SUSI) with targets selected according to its geographical
location. Narrow-angle astrometry using this indirect approach has achieved
sub-100 micro-arcsecond precision.
\end{abstract}


\section{Introduction}

Astrometry is an observation technique that measures the position of a celestial
object in the sky. It is not a new field of study but it has garnered recent
interest as a promising technique to search for extra-solar (exo-) planets
around main-sequence and intermediate-mass stars. The presence of an exoplanet
is inferred by observing the reflex motion of its host star due to gravitational
perturbation from the planet. In order to detect Jupiter-mass exoplanets,
astrometric precision in the regime of a few to tens of micro-arcseconds is
required to measure the motion, and hence the position of the candidate host
stars. At the current state-of-the-art, such a level of precision is only
attainable with relative astrometry (relative position with respect to a
reference in the sky) and with optical long baseline interferometry (OLBI) if
the observations are carried out from ground-based telescopes. OLBI is a
promising technique for ground-based high-precision narrow-angle astrometry
because the theoretical precision of the measurement is inversely proportional
to the baseline of the interferometer\cite{Shao:1992}. High-precision astrometry
has been successfully implemented with OLBI for exoplanets
search\cite{Muterspaugh:2010a} and currently there are several more attempts to
use this astrometric technique for various other field of
astronomy\cite{Delplancke:2008,Bartko:2009}.

There are at least three different approaches to measuring the relative position
of a star with OLBI. The first two approaches, directly and indirectly, measure
the difference in optical path length between the position of the central
fringes of the target and reference stars (or dOPD hereafter). In order to do
this, the differential phase of the interference fringes of the two stars, which
is proportional to the optical path difference per wavelength, is measured.  The
third approach measures the closure phase of the interference fringes of the two
stars (both the target and the reference stars are treated as one single
target). The first two approaches require a minimum number of one
interferometric baseline (or a set of two telescopes) to function while the
third approach requires the use of at least three baselines (or telescopes)
simultaneously. Using additional baselines with all approaches increases the
redundancy and the $uv$-coverage of an observation, therefore decreases the
amount of observation time needed for a given target. Among the approaches, the
direct differential phase measurement is the most commonly
implemented\cite{Muterspaugh:2010a,Delplancke:2008,Bartko:2009,Pott:2009,Kok:2013b}.

The direct measurement approach has been extensively discussed in the
literature. Instruments that implement this approach are engineered to measure
the optical path of the starlight to an accuracy of better than 5nm with a
baseline
in the regime of 100m in order to attain the required precision. Those
instruments typically employ a complex and usually costly network of laser
metrology systems for the optical path length
measurements\cite{Schuhler:2006,Gillessen:2012,Kok:2013}. Depending on the
circumstances of an observation, e.g.\ observing from a remote and poorly
accessible site in Antarctica which is thought to be the best site on Earth for
long baseline optical interferometry\cite{Tuthill:2012}, a complex optical and
mechanical design of an instrument can be a significant disadvantage.

On the contrary, the indirect approach to measuring dOPD, which is the subject
of this paper, may provide a simpler alternative solution.
It is a method that is widely used for measuring complex visibility of binary
stars but adapted for observing close binary stars which have separations wider
than the normal interferometric field of view.
Firstly, this paper discusses the basic principle and the limitation of the
indirect method. Then, it reports on some recent observations using such a
method carried out at the Sydney University Stellar Interferometer (SUSI) to
demonstrate the performance of the method. This paper also provides a suggestion
for a simple narrow-angle astrometric instrument that is suitable
for a variety of sites, including remote and poorly accessible observation site
such as Antarctica.

\section{Principle and limits}

The relative astrometry of two stars, $\Delta\vec{s}$, can be determined with an
interferometer of baseline, $\vec{B}$, from differential fringe phase, $\phi$,
and dOPD measurements because,
\begin{equation} \label{eq:pavo_phi} \begin{split}
  \phi &= 2\pi\sigma\,\text{dOPD}, \\
  \text{dOPD} &= \Delta\vec{s}\cdot\vec{B},
\end{split}\end{equation}
where $\sigma$ is the mean wavenumber\footnote{reciprocal of wavelength,
$1/\lambda$} of light at which the measurements were carried out. In the
indirect approach, $\phi$ and dOPD are inferred from the variation of the
visibility of the combined fringes of the target and the reference stars. The
visibility is higher when the fringes from the two stars are in phase (or when
$\phi$ is multiple of $2\pi$) and lower when the fringes are out of phase (or when
$\phi$ is multiple of $\pi$). The vector dot product in Eq.~\eqref{eq:pavo_phi}
varies with time as the pair of stars move across the sky due to the Earth's
rotation on its axis. As a result, the fringe visibility for a given pair of
stars has a characteristic sinusoidal modulation. The squared visibility of the
combined stellar fringes, $V^2$, derived from the van Cittert-Zernike theorem is
given as\cite{Tango:2006},
\begin{equation} \label{eq:pavo_v2bin} \begin{split}
V^2
= \frac{V_1^2 + V_2^2\beta^2\,\text{sinc}^2\left(\phi\,\delta\sigma/2\sigma\right) + 2\sqrt{V_1^2 V_2^2}\,\beta\,\text{sinc}\left(\phi\,\delta\sigma/2\sigma\right)\cos\phi}{\left(1+\beta\right)^2},
\end{split}\end{equation}
where $V_1^2$ and $V_2^2$ are the squared visibility of the fringes of the
individual stars, $\beta$ is the brightness ratio of the secondary star to the
primary, $\delta\sigma/\sigma$ is the bandwidth ratio of the spectral channel.
The \emph{sinc} function in the equation is defined as $\text{sinc}(x) = \sin\pi
x/\pi x$. The equation above assumed that the spectral response of the beam
combiner, the intensity of the source and the squared visibility of individual
stars are constant within the spectral channel the measurement is performed. By
measuring the fringe visibility of the two stars at different wavenumbers and
times, the dOPD and several other parameters ($\beta$,
$V_1^2$ and $V_2^2$) can be extracted by fitting the above model to the
measurements. Given the straightforward relation between dOPD and the
relative astrometry of the two stars, $\Delta\vec{s}$, it is common to extract
the latter directly from the model fitting instead. The uncertainty of the
estimated astrometry is proportional to the uncertainty of the fringe
visibility. The analytical form of the function, if it exists, is beyond the
scope of this paper to derive it.

Besides the straightforward data reduction from primary observables to relative
astrometry, the indirect approach method requires only one beam combining
instrument. As a result, it does not require any metrology systems to measure
non-common optical paths between beam combiners. This method is not entirely new
but has been used for very narrow-angle astrometry on spectroscopic binary
stars (whose component stars have on-sky separation of $\ll$500 milliarcseconds
(mas)) by several authors\cite{Armstrong:2004,Davis:2005,Tango:2009} and the
measurement
uncertainties obtained are in the sub-milliarcseconds regime. The work discussed
in this paper, however, attempted the indirect approach method on close binary
stars (whose component stars have on-sky separations of $<$2$''$).

There are three main challenges to be addressed when applying the indirect
approach method to close binary stars. Firstly, the beam combining instrument
must have a wide interferometric field of view. Secondly, observations of the
characteristic modulation of the fringe visibility can only be carried out
within a short time window at specific times. Thirdly, which is the most
challenging, unless the orientation of
the baseline used for observation is right for a given position angle of a
binary star the observation time window may not exist at all. The first two
challenges essentially set the performance criteria for the intended beam
combining instrument.

The field of view of a beam combiner determines the largest star separation that
the instrument can observe. It varies with design but is usually very
narrow, e.g.\ $\lesssim$500mas for NPOI classic\cite{Armstrong:2004}, MIRC at
CHARA\cite{Monnier:2004} and AMBER at the Very Large Telescope
Interferometer\cite{Petrov:2007} (VLTI). This is one reason why targets for past
experiments are limited to binaries of very
small separations. On the contrary, the PAVO beam combining instrument at SUSI
which is used to demonstrate the indirect approach method has a wider field of
view. Its field of view of $<$2$''$ enables PAVO to observe binaries with larger
separations than previously attempted. The optical design of PAVO has been
described in various other papers\cite{Ireland:2008,Robertson:2010}
and therefore is not discussed in detail in this paper.

The time window when the fringe visibility modulation can be observed depends on
the coherence length of the fringes and the rate of change of dOPD between
fringes of the two stars. During this time window, the fringe packets of the two
stars crossover each other. The term fringe packet describes the localization of
interference fringes within the coherence length due `bandwidth smearing'.  The
visibility modulation diminishes as dOPD becomes larger than the coherence
length and this effect is modeled by the \emph{sinc} function term in
Eq.~\eqref{eq:pavo_v2bin}. In a practical beam combiner, the coherence length of
stellar fringes is in the order of several tens of micrometers ($\mu$m). The
longer the coherence length the longer the observation time window is. With a
simple North-South (N-S) baseline at latitude similar to SUSI, the rate of
change of dOPD, which is proportional to the binary star separation, can reach
up to tens of nanometers per second per arcsecond per 100m of baseline.
Therefore, the window of opportunity to observe the visibility modulation of
binary of a given separation with a
100m baseline has an upper limit of about tens of minutes. This limit is adequate for
averaging out the astrometric bias in the differential phase measurement due to
anisoplanatism in the turbulent atmosphere\cite{Shao:1992}, even at an average
astronomical site like SUSI. Nevertheless, a longer integration time is always
advantageous for astrometry.

Other than that, the fast changing fringe visibility during the observation time
window also gives rise to another requirement for the beam combining instrument
that is to
be used. The total integration time per $V^2$ measurement, $T_{\text{int}}$,
must be much smaller than the time for dOPD to change by one wavelength,
$T_{\text{int}} \ll \left(\pi\,\delta\sigma\,d\text{dOPD}/dt\right)^{-1}$.
The factor should be lesser than half in order to keep the systematic error of the
$V^2$ measurements much smaller than its peak-to-peak variation. The limit for
$T_{\text{int}}$ ranges from several seconds for observations in the visible
wavelengths to several tens of seconds for observations in the near infra-red
(IR) wavelengths. In general, beam combiners make a single $V^2$ measurement
every several to tens of milliseconds but incoherently integrate several
measurements together to increase signal-to-noise ratio (SNR). By limiting the
total integration time, the SNR of the $V^2$ data points used for the model
fitting is also limited. This eventually limits the astrometric precision due to
random errors that can be achieve with the model fitting. So we need a sensitive
beam combiner so that precise measurement of $V^2$ can be done in a short time.

The fringe packet crossover event, described in the previous paragraph, occurs
when
the line joining the two stars of the binary
is perpendicular to
the baseline, i.e.\ when the $\Delta\vec{s}\cdot\vec{B}$ term in
Eq.~\eqref{eq:pavo_phi} is zero. This position-angle-baseline alignment does not
always occur when a targeted binary star is observable. A target is considered
observable when it is at a high enough elevation in the sky after sunset and
before sunrise. For example, only $\sim$20 out of $\sim$70 close binary stars
brighter than magnitude 5 have fringe packet crossover events that are observable
with a N-S baseline at SUSI and at the VLTI (baselines H0-J4). The number of stars in
the example was determined by querying the Washington Double Star (WDS) catalog.
The details of the query are given in the next section. The number of stars
observable with this method can certainly be increased by adding more baselines
of different orientations. For example, the number of binary stars exhibiting
the fringe packet crossover event increases to $\sim$40 out of $\sim$70 if an
additional East-West (E0-J1) baseline is included at the VLTI. It would take an
uneconomical number of baselines, or telescopes, at a mid-latitude site to be
able to always observe the fringe packet crossing event of any binary star at
an arbitrary position angle. This could be one of the reasons why the indirect
approach is less popular for narrow-angle astrometry. On the contrary, this
approach could be favorable for observations from Antarctica because such an
event can be observed for every observable binary star with just one baseline
because the sun does not rise during the Southern Hemisphere winter. Therefore,
there are two observable fringe crossover events per sidereal day for each
binary star in Antarctica.

\section{Observations}

Table~\ref{tab:targets} compares the number of binary stars that exhibit fringe
crossover events at various observation sites using a N-S baseline. Also listed
in the table is the total number of binary stars observable from those sites.
The targets are queried from the WDS catalog with the following criteria.
\begin{itemize}
  \item Magnitude of the primary component, $m_V < 5$
  \item Difference of magnitude between components, $\Delta m < 2$
  \item Binary star separation, $\rho < 2''$
  \item Zenith distance during transit, $\left|\delta - \phi_{\text{LAT}}\right| < 60^{\circ}$
\end{itemize}
The total number of targets observable is systematically lower at Dome C,
Antarctica compared to the other two sites due to the distribution of bright
binary stars over the Southern hemisphere sky. However, unlike in mid-latitude
sites, all targets have two observable fringe crossover events per day in
Antarctica.

\begin{table}
\center
\caption{Observable targets$^{\dagger}$ with various long baseline interferometers.}
\label{tab:targets}
\begin{tabular}{@{\hspace{0.7cm}}c@{\hspace{0.7cm}} @{\hspace{0.7cm}}c@{\hspace{0.7cm}} @{\hspace{0.7cm}}c@{\hspace{0.7cm}}}
\toprule
SUSI & VLTI & Dome C$^{\#}$ \\
\midrule
\multirow{2}{*}{20 (80)} & 16 (68)     & \multirow{2}{*}{39 (39)} \\
			 & 42$^*$ (68) &                          \\
\bottomrule
\multicolumn{3}{l}{\footnotesize{$^{\dagger}$with (and without) events using a N-S baseline}} \\
\multicolumn{3}{l}{\footnotesize{$^*$using an addtional E-W baseline}} \\
\multicolumn{3}{l}{\footnotesize{$^{\#}$site used as a proxy for an interferometer in Antarctica}} \\
\end{tabular}
\end{table}

\begin{figure}
\centering
\includegraphics[width=0.8\textwidth]{\imgdir/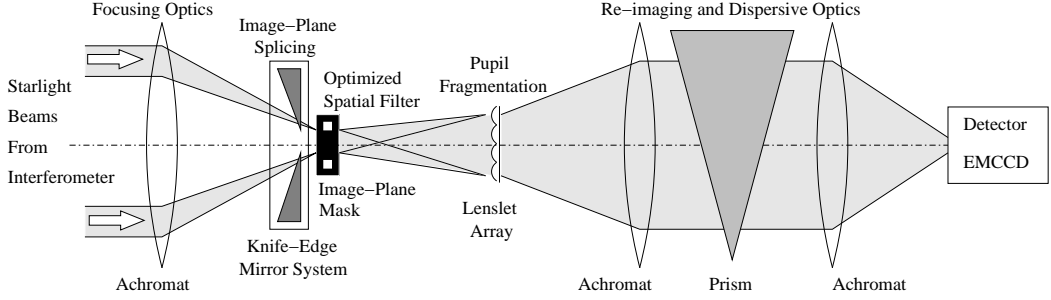}
\vspace{1em}
\caption[]{Schematic diagrom of the PAVO beam combiner at SUSI.}
\label{fig:opt_pavo_sch}
\end{figure}

In order to demonstrate the performance of the indirect approach to narrow-angle
astrometry of close binary stars, observations of two targets ($\gamma$~Lupi and
$\zeta$~Sagittarii) were carried out at SUSI using the PAVO beam combiner.  The
PAVO beam combiner was assembled and obtained its first stellar fringes in
November 2008. A similar setup\cite{Ireland:2008} is also operational at the
Center for High Angular Resolution Astronomy (CHARA) array. PAVO is a
multi-axially aligned Fizeau-type interferometer. But unlike a typical Fizeau
interferometer, PAVO forms spatially modulated interference fringes in the pupil
plane of the interferometer and then spectrally disperses the fringes with a low
resolution (R$\sim$50) spectrograph. It also employs spatial filtering in its
image plane with two $\sim$1.2mm square apertures and an array of cylindrical
lenslets to utilize the full multi-r$_0$ aperture of the siderostats at SUSI.
The interferometric field of view of PAVO at SUSI is $<$2$''$.
Fig.~\ref{fig:opt_pavo_sch} shows the schematic diagram of the PAVO beam
combiner. 

\subsection{$\gamma$~Lupi}

$\gamma$~Lupi (HR5776; Gam Lup) is a triple star
system where its primary component ($\gamma$~Lupi~A) is itself a suspected
spectroscopic binary\cite{Levato:1987,Docobo:2006} ($\gamma$~Lupi~Aa-Ab). Its primary and
secondary components ($\gamma$~Lupi~A-B) form a visual binary.
Table~\ref{tab:obs_gamlup} shows the successful observations of
$\gamma$~Lupi~A-B carried out with PAVO and the calibrator stars
used for the data analysis.

\begin{table}
\centering
\caption{Successful observations of $\gamma$ Lupi A-B}
\label{tab:obs_gamlup}
\begin{tabular}{c c c l}
\toprule
Date & Baseline & Range of HAs (Hr) & Calibrators \\
\midrule
100805 & N3-S1 & 0.99 -- 1.50 & HR4205, Tet~Lup \\
100806 & N3-S1 & 1.02 -- 1.35 & 13~Sco \\
100813 & N3-S3 & 0.74 -- 1.02 & Alf~Tel \\
130705 & N4-S2 & 0.60 -- 1.20 & Bet~Lup, Del~Lup \\
\bottomrule
\multicolumn{4}{l}{\footnotesize{N3-S1=15m, N3-S3=40, N4-S2=60m}} \\
\end{tabular}
\end{table}

The full description of the data reduction pipeline from PAVO interferograms to
calibrated $V^2$ data is beyond the scope of this paper. The pipeline is similar
to the one used for reducing interferograms of the PAVO at CHARA beam
combiner\cite{Maestro:2012}.

Each set of calibrated $V^2$ data (one set per successful observation) was used
to fit a binary star model (see Eq.~\eqref{eq:pavo_v2bin}) and to extract the
relative astrometry of the binary star. The relative astrometry of a binary star
is expressed in a polar coordinate system where the radial axis, $\rho$,
measures the angular separation between the secondary and the primary component
while the position angle axis, $\theta$, measures the bearing of the secondary
component with respect to the celestial North. The 2D plane which the coordinate
system defines is tangential to the celestial sphere. The relation between
$\rho$ and $\theta$ with the equatorial coordinate system is,
\begin{equation}
  \Delta\alpha = \rho \sin \theta,\quad
  \Delta\delta = \rho \cos \theta,
\end{equation}
where $\Delta\alpha$ and $\Delta\delta$ are the small angle approximation of
relative astrometry in the right ascension and declination axes respectively.

\begin{figure}
\centering
\subfloat[]{\includegraphics[width=0.75\textwidth]{\imgdir/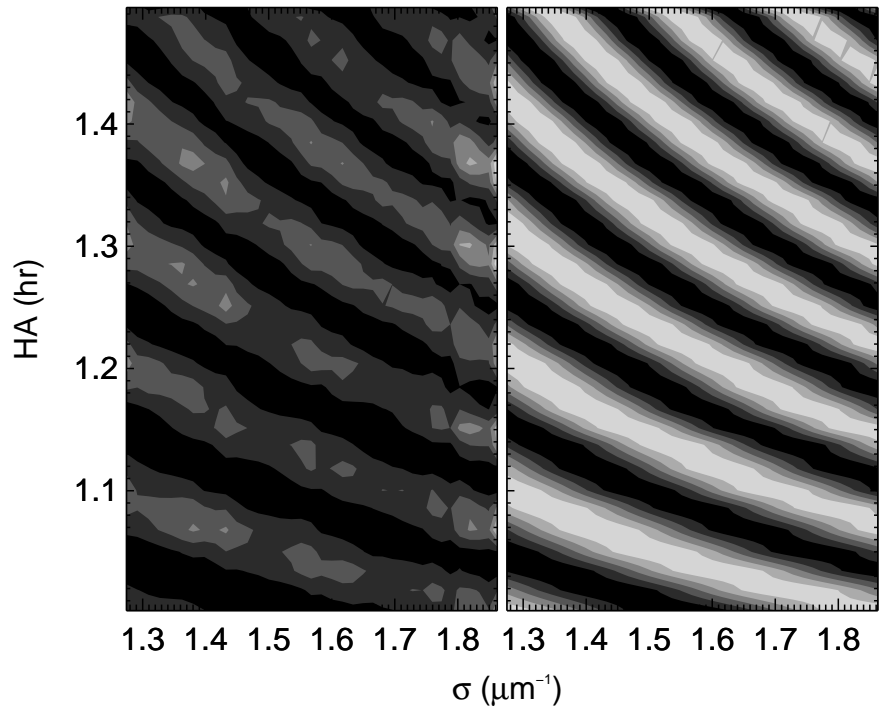}}\\
\vspace{-1.5em}
\subfloat[]{\includegraphics[width=0.75\textwidth]{\imgdir/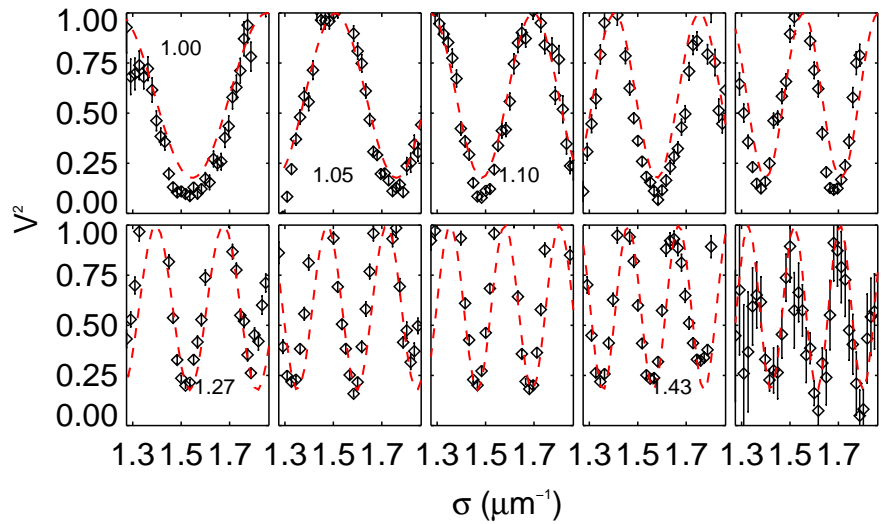}}
\caption[]{Calibrated $V^2$ (left panel of (a)) of $\gamma$~Lupi~A-B obtained with PAVO on 5th August 2010 and the best fit binary star model (right panel of (a)). The grayscale in the images in (a) represents the relative amplitude of $V^2$ within each image. The difference in contrast between the data and model does not represent a discrepancy. Cross-sections of (a) at different hour angles (indicated in the legend) are plotted in (b) as $V^2$ versus wavenumber with the best fit model represented by the dashed lines.}
\label{fig:obs_gamlup_100805}
\end{figure}

\begin{figure}
\centering
\subfloat{\includegraphics[width=0.45\textwidth]{\imgdir/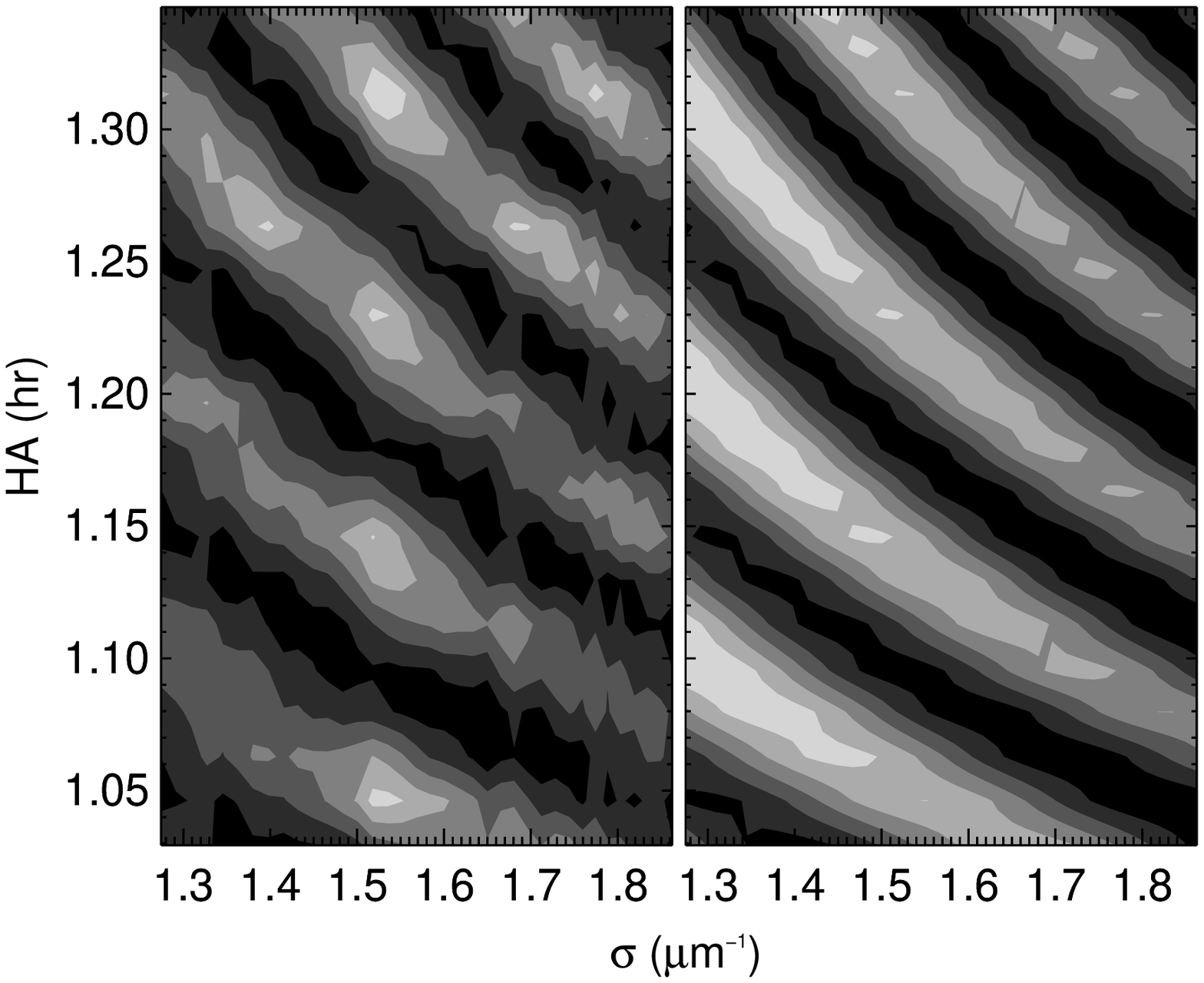}}
\subfloat{\includegraphics[width=0.55\textwidth]{\imgdir/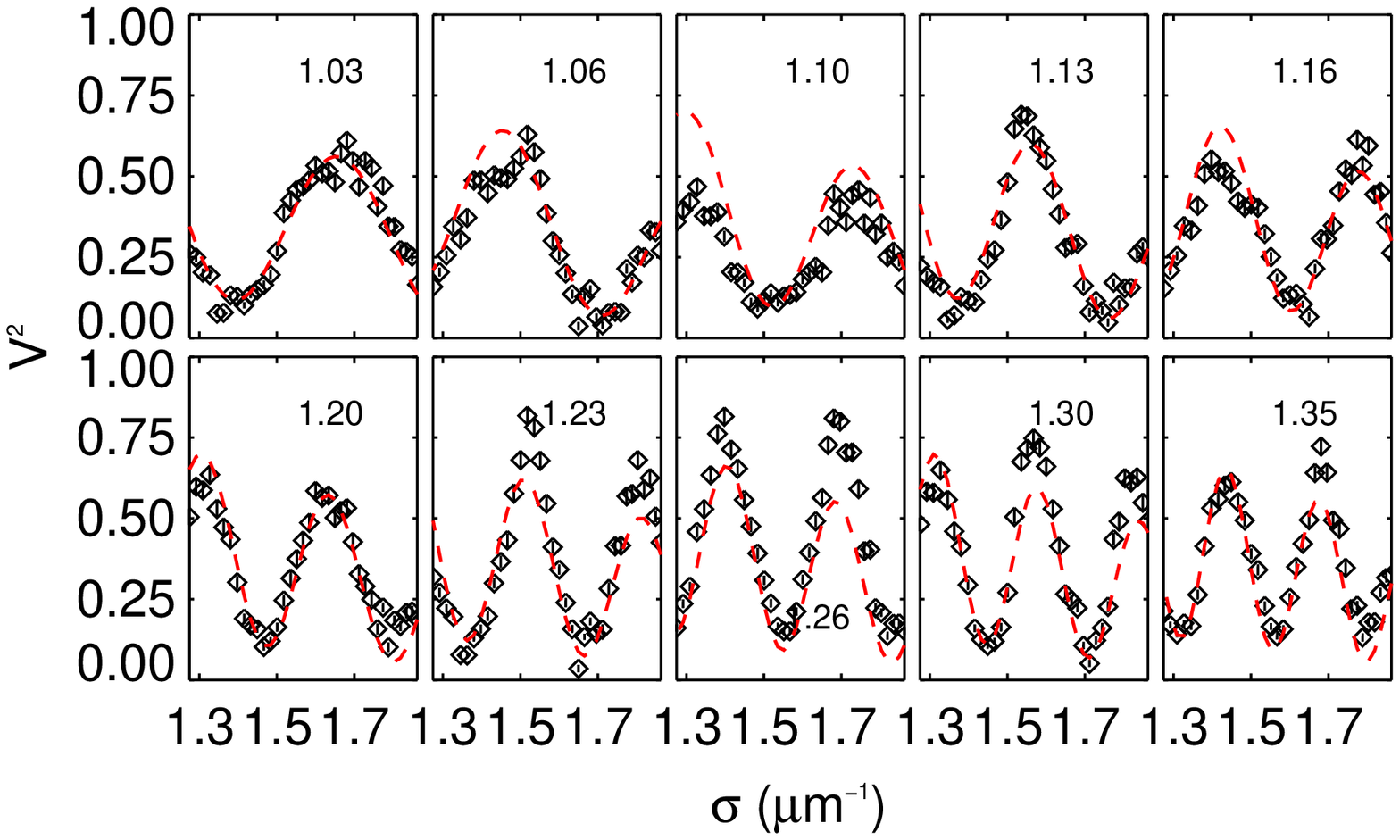}}\\
\vspace{-1em}
\subfloat{\includegraphics[width=0.45\textwidth]{\imgdir/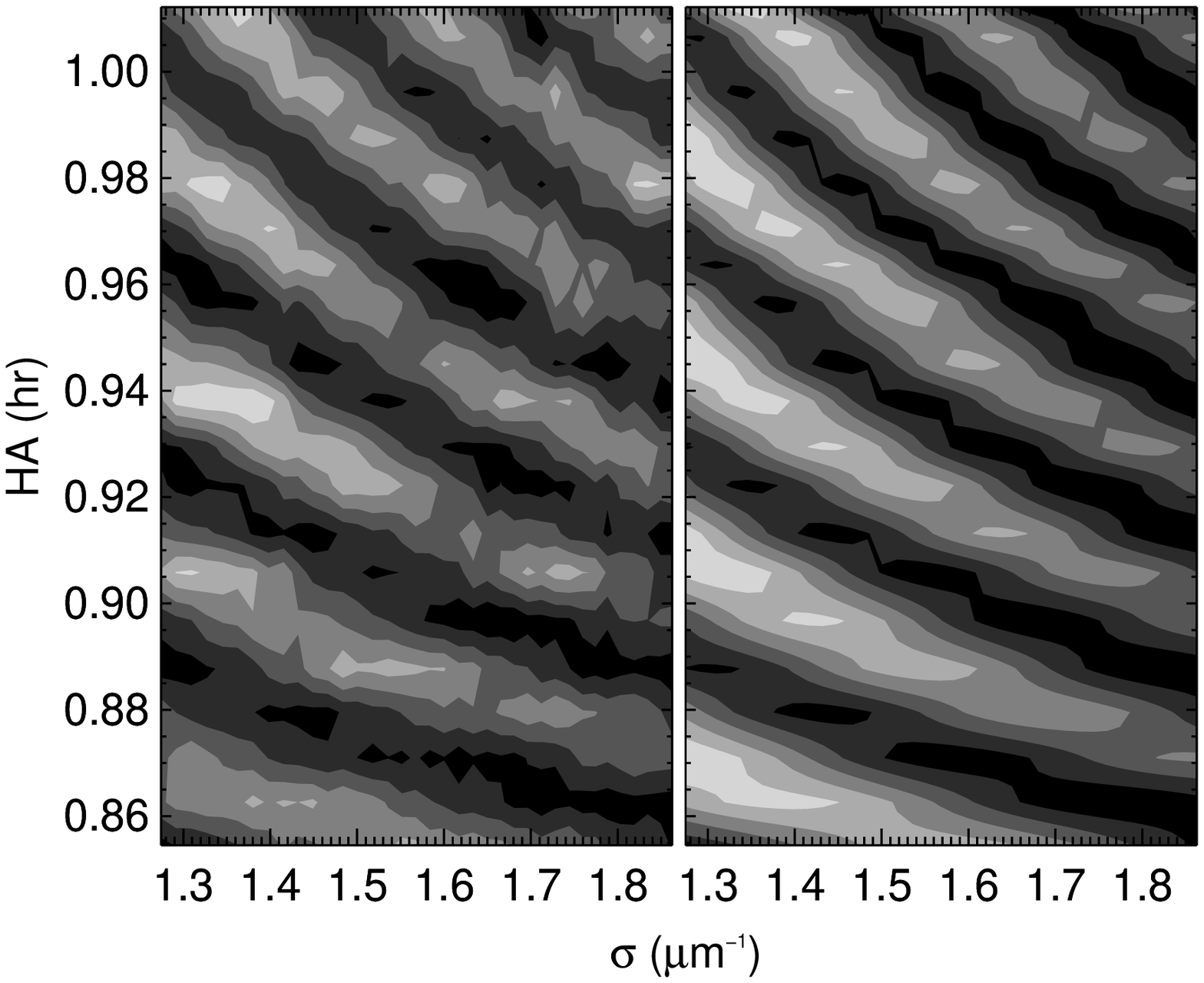}}
\subfloat{\includegraphics[width=0.55\textwidth]{\imgdir/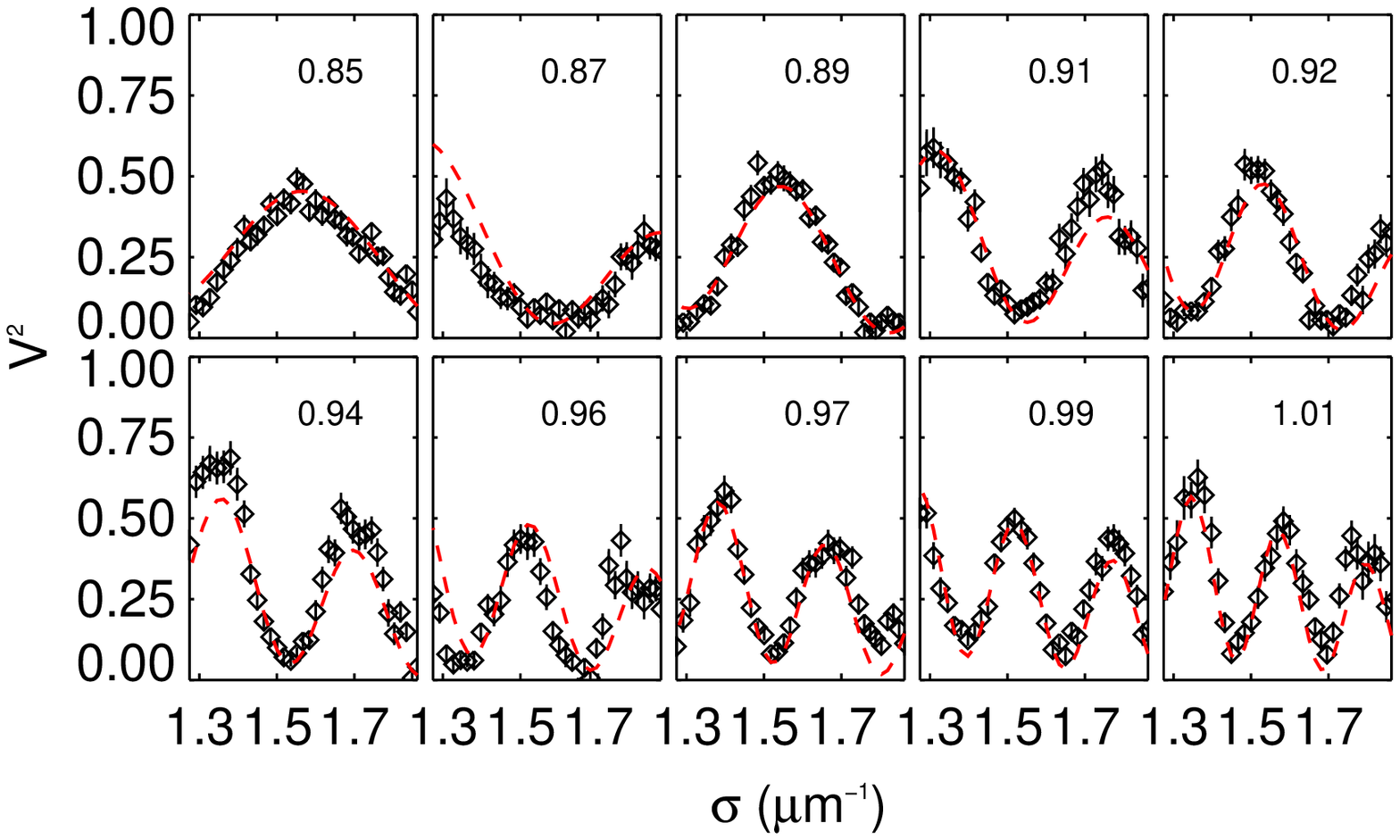}}\\
\vspace{-1em}
\subfloat{\includegraphics[width=0.45\textwidth]{\imgdir/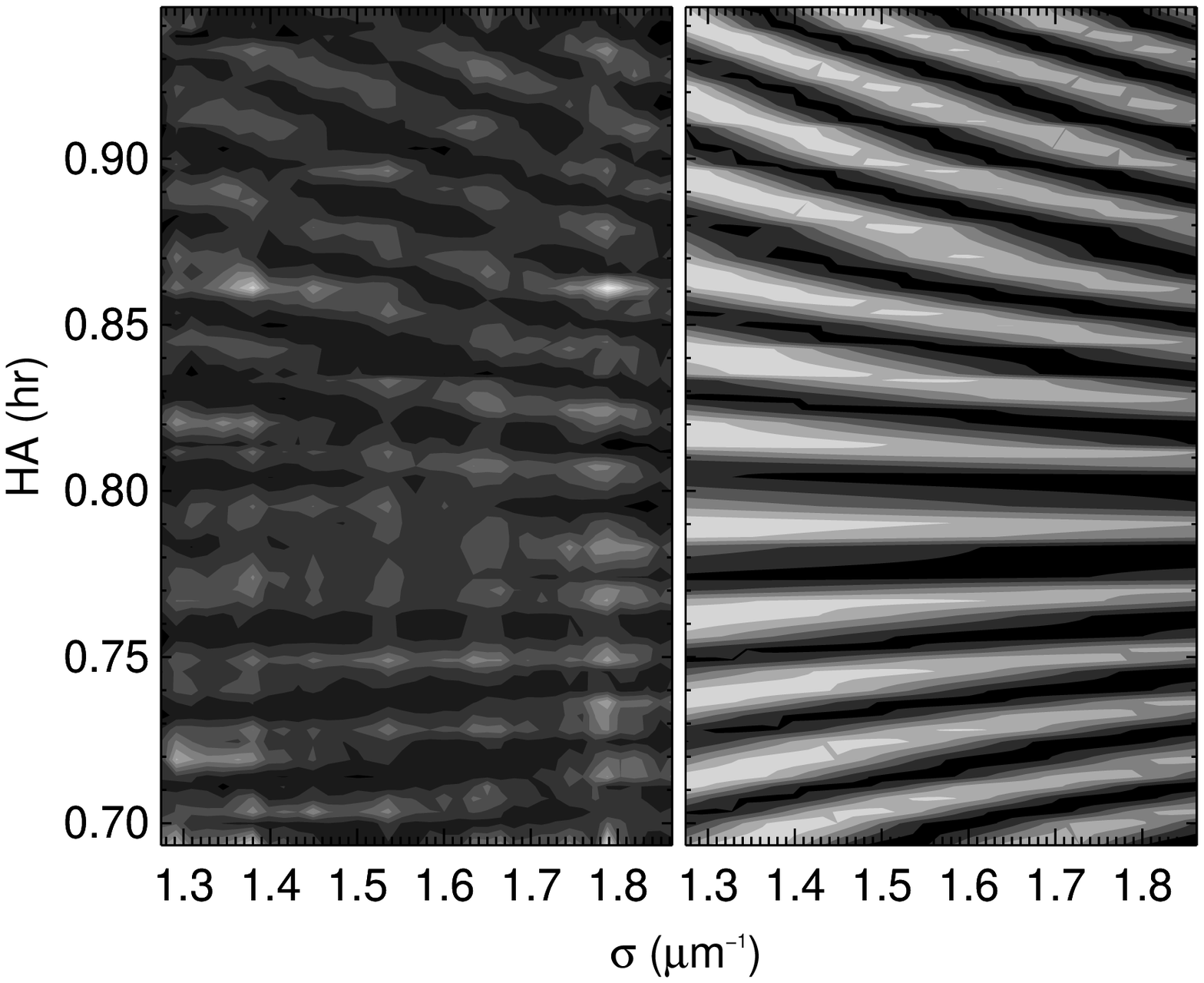}}
\subfloat{\includegraphics[width=0.55\textwidth]{\imgdir/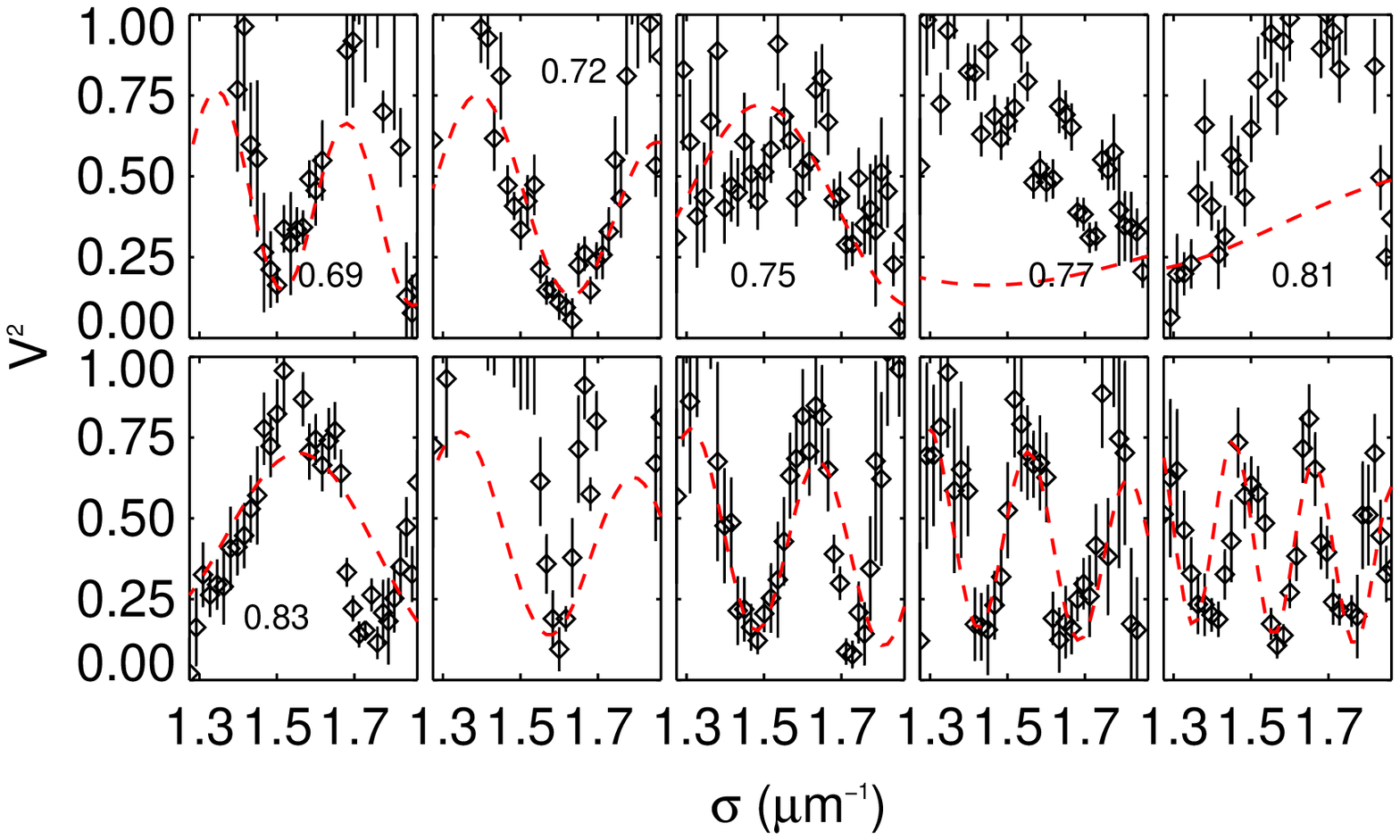}}\\
\caption[]{Similar to Fig.~\ref{fig:obs_gamlup_100805} but with data taken on 6th \& 13th August 2010 and 5th July 2013 for plots on each row from top to bottom.}
\label{fig:obs_gamlup_1xxxxx}
\end{figure}

Comparisons between the two-dimensional $V^2$ data and the fitted model are
shown in Fig.~\ref{fig:obs_gamlup_100805}--\ref{fig:obs_gamlup_1xxxxx} in the
form of grayscale images. The grayscale level represents the relative amplitude
of $V^2$ in a given image.
A fringe crossover event was successfully observed in
The last image in Fig.~\ref{fig:obs_gamlup_1xxxxx} shows the climax of a fringe
crossover event at HA$\sim$0.8 when the $V^2$ modulation stripe pattern is
perpendicular to the HA axis. Despite missing the climax, the fringe crossover
event was still in progress in all the other images.
Also shown in the figures are cross-sectional
plots of $V^2$ versus wavenumber at different hour angles.
In the model fitting, the parameter $V^2_1$ was allowed to vary but $V^2_2$ was
kept at 1 because the primary component of the binary is suspected to have a
spectroscopic companion which may reduce fringe visibility while the secondary
is estimated to be unresolved (UD$\lesssim$0.5mas).
The accuracy of the values of $\beta$, $V^2_1$ and $V^2_2$ are less important as
far as astrometry is concerned because the sinusoidal component of the
visibility modulation depends only on the term $\phi$ in Eq.~\eqref{eq:pavo_phi}
which is related to the fringe packet separation of the primary and secondary
components of the binary.
Furthermore, the values of the 3 parameters are sensitive to how well the
$V^2$ data are calibrated.

One drawback of this approach to determine the fringe packet separation
indirectly from the modulation of the square of the fringe visibility is that
the estimated position angle has a 180$^\circ$ ambiguity. This ambiguity arises
because the position of individual fringe packets could not be determined from
the visibility modulation. Therefore the position angle extracted from the model
fit could either be 96$^\circ$ or 276$^\circ$, if without any prior knowledge.
In the case of $\gamma$~Lupi, the ambiguity is resolved by cross checking the
values obtained with other techniques (e.g.\ speckle
interferometry\cite{Tokovinin:2010}).

Table~\ref{tab:results_gamlup} shows the parameters extracted from the
model-fitting as well as the number of $V^2$ data used and the reduced $\chi^2$
of the each fit. The values of reduced $\chi^2$ are larger than unity because a
binary model is used in the model-fitting despite the primary component having a
suspected spectroscopic companion. The discrepancy between the model and data is
most obvious (see Fig.~\ref{fig:obs_gamlup_1xxxxx} when HA$\approx$0.79) at a
time close to the instance when the fringe packets cross over because the
modulation of the fringe visibility is dominated by the astrophysical feature of
the primary component. Nevertheless, the astrometry of the wider pair
($\gamma$~Lupi~A-B) based on the fitting of the sinusoidal component of the
$V^2$ model is still good. The uncertainty numbers in the table are not scaled
with $\sqrt{\chi^2_R}$ and do not contain systematic errors which are discussed
in the next paragraph.

\begin{table}
\centering
\caption{Relative astrometry of $\gamma$ Lupi A-B}
\label{tab:results_gamlup}
\begin{tabular}{l r r r r}
\toprule
Parameter &
2010.5925 &
2010.5952 &
2010.6143 &
2013.5078 \\
\midrule

$N_{V^2}$ &
37$\times$31 &
37$\times$20 &
37$\times$31 &
37$\times$142 \\

LST$_X$ (Hr) &
16.371 &
16.376 &
16.378 &
16.374 \\

$\beta_{\rm{A,B}}$ &
0.408$\pm$0.001 &
0.312$\pm$0.001 &
0.304$\pm$0.003 &
0.303$\pm$0.004 \\

$\Delta\alpha_{\rm{A,B}}$ (mas)	&
-798.40$\pm$0.26 &
-803.64$\pm$0.29 &
-805.06$\pm$0.59 &
-811.81$\pm$0.31 \\

$\Delta\delta_{\rm{A,B}}$ (mas)	&
84.76$\pm$0.04 &
85.82$\pm$0.05 &
85.86$\pm$0.07 &
86.05$\pm$0.03 \\

$\rho_{\rm{A,B}}$ (mas) &
802.9$\pm$0.3 &
808.2$\pm$0.3 &
809.6$\pm$0.6 &
816.4$\pm$0.3 \\

$\theta_{\rm{A,B}}$ ($^\circ$) &
276.05996$\pm$0.00002 &
276.09546$\pm$0.00003 &
276.08760$\pm$0.00008 &
276.05063$\pm$0.00001 \\

$\chi^2_{\rm{R}}$ &
32.0 &
43.5 &
 4.4 &
 9.1 \\

\bottomrule
\multicolumn{5}{l}{\footnotesize{Systematic errors are not included in the
uncertainties.}} \\
\end{tabular}
\end{table}

\begin{figure}
\centering
\subfloat[]{\includegraphics[width=0.35\textwidth]{\imgdir/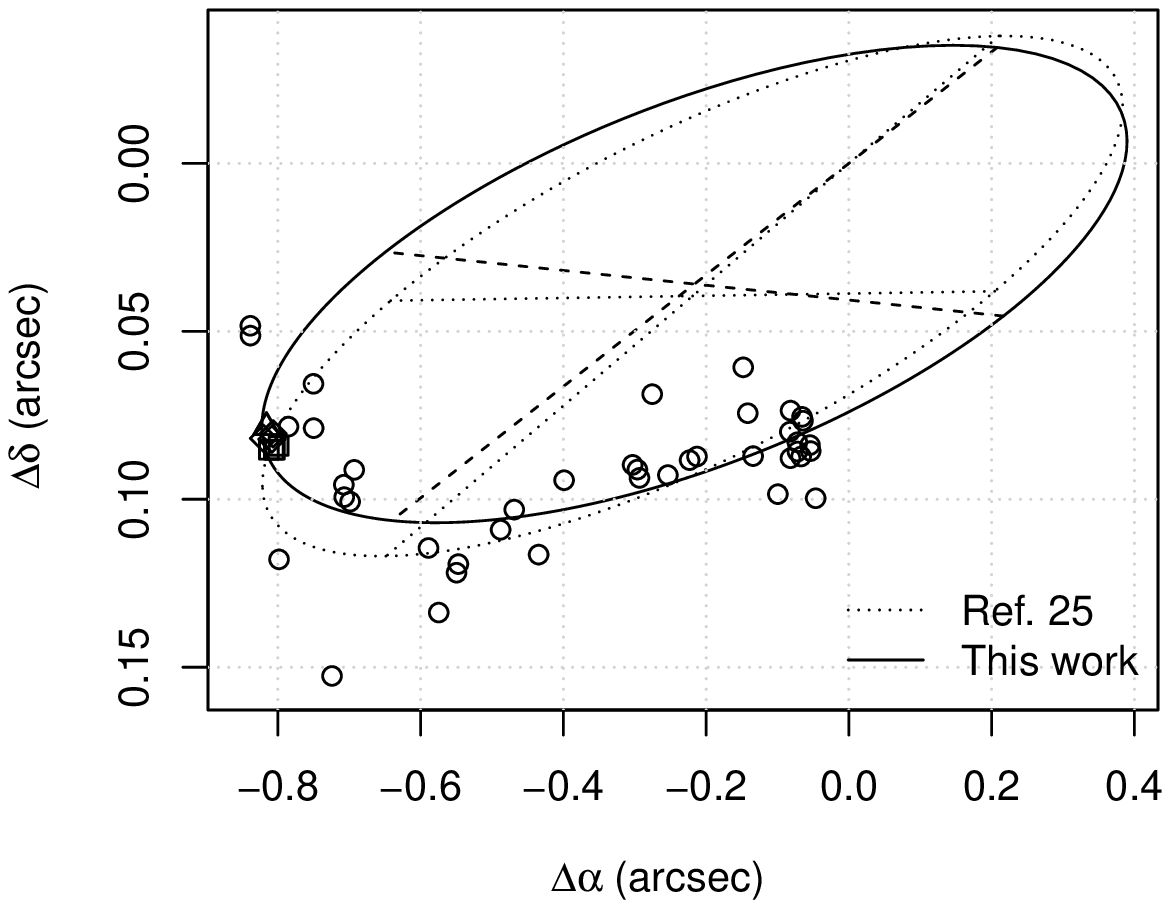}}
\subfloat[]{\includegraphics[width=0.63\textwidth]{\imgdir/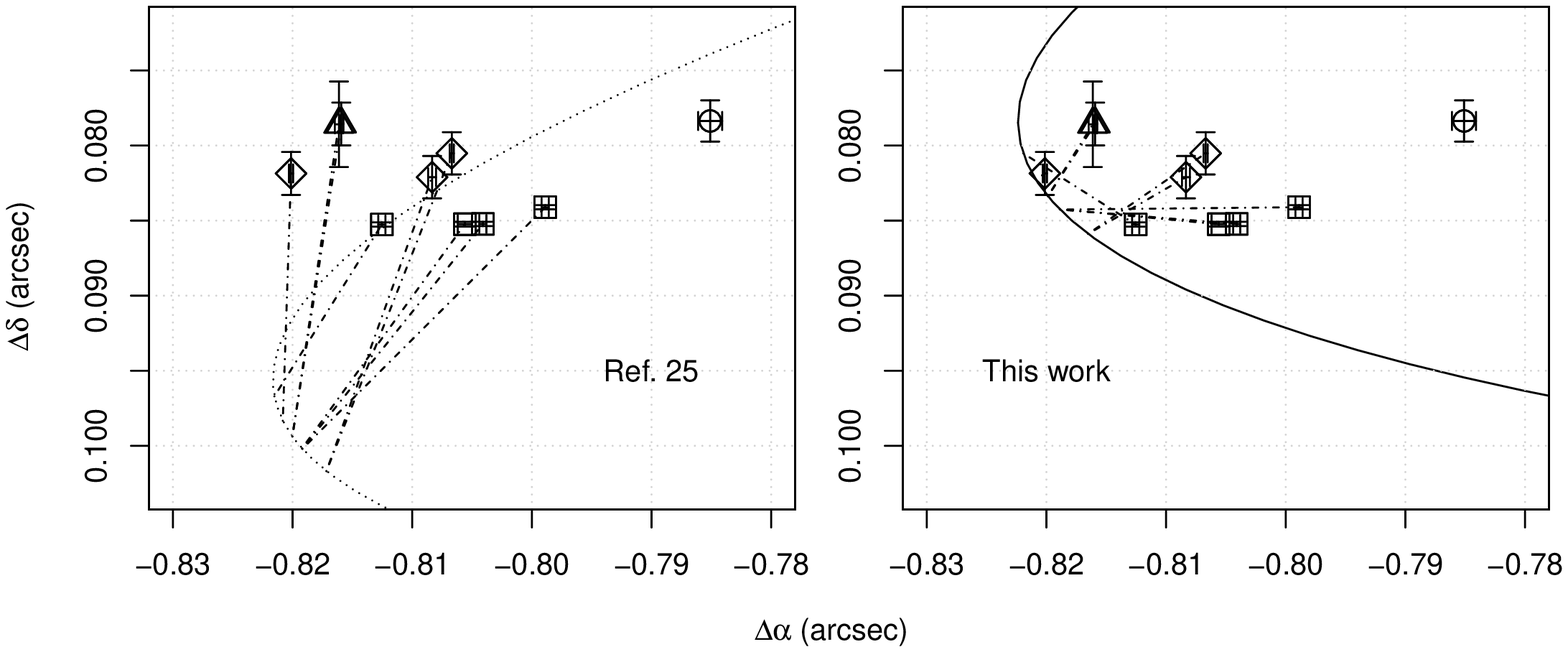}}
\caption[]{
  The relative positions of $\gamma$~Lupi~A-B. The circular ($\bigcirc$) points
  are data obtained from the WDS (1835--1988) catalog and the Fourth Catalog of
  Interferometric Measurements of Binary Stars (1933--2006); the diamond
  ($\Diamond$) points are from Ref.~\citenum{Tokovinin:2010} and
  Ref.~\citenum{Tokovinin:2012}; the triangular ($\triangle$) points are from
  Ref.~\citenum{Hartkopf:2012}; and the square ($\Box$) points are from this
  work.
  The ellipses in (a) and partially in (b) describe the
  orbit of the secondary component around the primary, which is centered at the
  origin in the plots. The orbital elements for
  the dotted elliptical lines are obtained from Ref.~\citenum{Heintz:1990}
  while the solid elliptical lines are refitted with the available data. The
  period, $P$ ($=190$), the epoch of periastron, $T$ ($=1885.0$), and the
  eccentricity, $e$ ($=0.51$), of the orbit were kept the same. The updated
  values for the semimajor axis, $a$, the longitude of periastron, $\omega$, the
  longitude of ascending node, $\Omega$, and the inclination, $i$, of the orbit
  are 0.659$''$, 310.4$^\circ$, 93.2$^\circ$ and 95.3$^\circ$ respectively.
}
\label{fig:orb_hr5776}
\end{figure}

Table~\ref{tab:results_gamlup_j2000} shows the same extracted parameters but
expressed in an equinox 2000 equatorial coordinate system. Also given in the
table are astrometry from other techniques (mainly speckle interferometry) for
comparison. The values
reported in Ref.~\citenum{Tokovinin:2010} in a series of papers given in the
table have systematic errors of $\lesssim$0.5\% in binary separation and
$\sim$2$^\circ$ in
position angle\cite{Hartkopf:2012}. Similarly, due to uncertainties in the
baseline solutions and the wavelength scale used in the data reduction, the
binary separation and position angle measurements obtained with PAVO may have
systematic errors not reflected in the uncertainty numbers reported in the
table. For most of the PAVO observations, instead of using a more precise approach
\cite{Kok:2013b}, optical alignment was carried out with the standard SUSI
alignment procedure. The latter approach typically aligns the apertures of the
PAVO 2-hole mask with the pivot points of the siderostats to within 2cm.
Assuming this offset is parallel to the
$uv$-plane centered on the primary component of the binary star (the worst
case), this translates to a systematic error of $<$0.08$^\circ$ (upper limit set
by the shortest 15m
baseline) in the position angle measurements. An uncertainty in the baseline
solution also translates to an uncertainty in the binary separation but the
magnitude is smaller than the effect of an uncertainty in the wavelength scale.
The estimated relative uncertainty in the wavelength scale is
$\sim$1\% ($\sim$6nm over $\sim$0.6$\mu$m). This estimation is based on the
pixel offset value obtained during the PAVO wavelength scale calibration, which
is typically $\lesssim$1 pixel or half of the width of one spectral channel.
Since the relative uncertainty in wavelength scale directly translates to an
uncertainty in the projected fringe
packet separation measurement the systematic error in the binary separation
measurements for $\gamma$~Lupi~A-B is $\sim$8mas ($\sim$1\% of $\sim$0.8$''$).
Therefore the measurements obtained with PAVO are consistent with the values
obtained with speckle
interferometry\cite{Tokovinin:2010,Tokovinin:2012,Hartkopf:2012}. The relative
astrometry of the binary from various sources and the best fitted orbit are
plotted in Fig.~\ref{fig:orb_hr5776} for comparison.

\begin{table}
\centering
\caption{Comparison of $\gamma$~Lupi~A-B astrometry.}
\label{tab:results_gamlup_j2000}
\begin{tabular}{l l l l l l}
\toprule
Source & Julian yr. & $\theta$   & $\delta\theta$ & $\rho$ & $\delta\rho$ \\
       & (+2000)    & ($^\circ$) & ($^\circ$)     & ($''$) & (mas) \\
\midrule
\multirow{2}{*}{TOK2010$^\dagger$}
       & 09.2603$^y$ & 275.7  & 0.0  & 0.8107 & 0.1 \\   
       & 09.2603$^H$ & 275.8  & 0.1  & 0.8125 & 0.3 \\   
 & & & & & \\
\multirow{4}{*}{This work$^{**}$}
       & 10.5925     & 276.01 & 0.00 & 0.8033 & 0.3 \\
       & 10.5952     & 276.05 & 0.00 & 0.8086 & 0.3 \\
       & 10.6143     & 276.04 & 0.00 & 0.8101 & 0.6 \\
       & 13.5078     & 275.99 & 0.00 & 0.8170 & 0.3 \\
 & & & & & \\
\multirow{2}{*}{HAR2012$^\#$}
       & 11.3028$^y$ & 275.5  & 0.2  & 0.8199 & 0.2 \\   
       & 11.3028$^I$ & 275.5  & 0.0  & 0.8197 & 0.0 \\   
 & & & & & \\
TOK2012$^\ddagger$
       & 12.1845$^y$ & 275.7  & 0.0  & 0.8242 & 0.1 \\   
Ephemeris$^*$
       & 13.5000     & 276.7  & --   & 0.8271 & --  \\   
\bottomrule
\multicolumn{6}{l}{\footnotesize{$^\dagger$Ref.~\citenum{Tokovinin:2010}}} \\
\multicolumn{6}{l}{\footnotesize{$^{**}$see Table~\ref{tab:results_gamlup} for $\delta\theta$}} \\
\multicolumn{6}{l}{\footnotesize{$^\#$Ref.~\citenum{Hartkopf:2012}}} \\
\multicolumn{6}{l}{\footnotesize{$^\ddagger$Ref.~\citenum{Tokovinin:2012}}} \\
\multicolumn{6}{l}{\footnotesize{$^*$based on a grade 3 orbit from Ref.~\citenum{Heintz:1990}}} \\
\multicolumn{6}{l}{\footnotesize{$^y$measured at $\sim$550nm wavelength}} \\
\multicolumn{6}{l}{\footnotesize{$^H$measured at H$\alpha$ ($\sim$650nm) wavelength}} \\
\multicolumn{6}{l}{\footnotesize{$^I$measured at $\sim$770nm wavelength}} \\
\end{tabular}
\end{table}

\subsection{$\zeta$~Sagittarii}

$\zeta$~Sagittarii (HR7194; Zet Sgr) is a close binary star system which has an
expected on-sky separation of 0.5$''$, a position angle of 265$^\circ$ and an
orbital period of 21 years\cite{De-Rosa:2012}. The primary component
($\zeta$~Sagittarii~A) is suspected to have an unresolved companion based on a
discrepancy between the observed dynamical mass of the binary and its
theoretical mass estimated from mass-magnitude relation\cite{De-Rosa:2012}.

Table~\ref{tab:obs_zetsgr} shows the number of observations attempted on this
binary system. Despite being unsuccessful, the observation in 2012 hinted at the
presence of a tertiary component because the PAVO $V^2$ data did not show any
modulation and the fringe visibility was low, even with a short baseline. A
successful follow up observation in 2013 resolved the tertiary component
($\zeta$~Sagittarii~Ab).

\begin{table}
\centering
\caption{Successful and failed observations of $\zeta$ Sagittarii}
\label{tab:obs_zetsgr}
\begin{tabular}{c c l c l}
\toprule
Date & Baseline & Outcome & Range of HAs (Hr) & Calibrators \\
\midrule
121003 & N3-S1 & Fail    &  2.5 -- 3.5  & Phi~Sgr \\
130726 & N4-S2 & Success & -1.0 -- -0.3 & Phi~Sgr \\
\bottomrule
\multicolumn{4}{l}{\footnotesize{N3-S1=15m, N4-S2=60m}} \\
\end{tabular}
\end{table}

\begin{figure}
\centering
\subfloat[]{\includegraphics[width=0.5\textwidth]{\imgdir/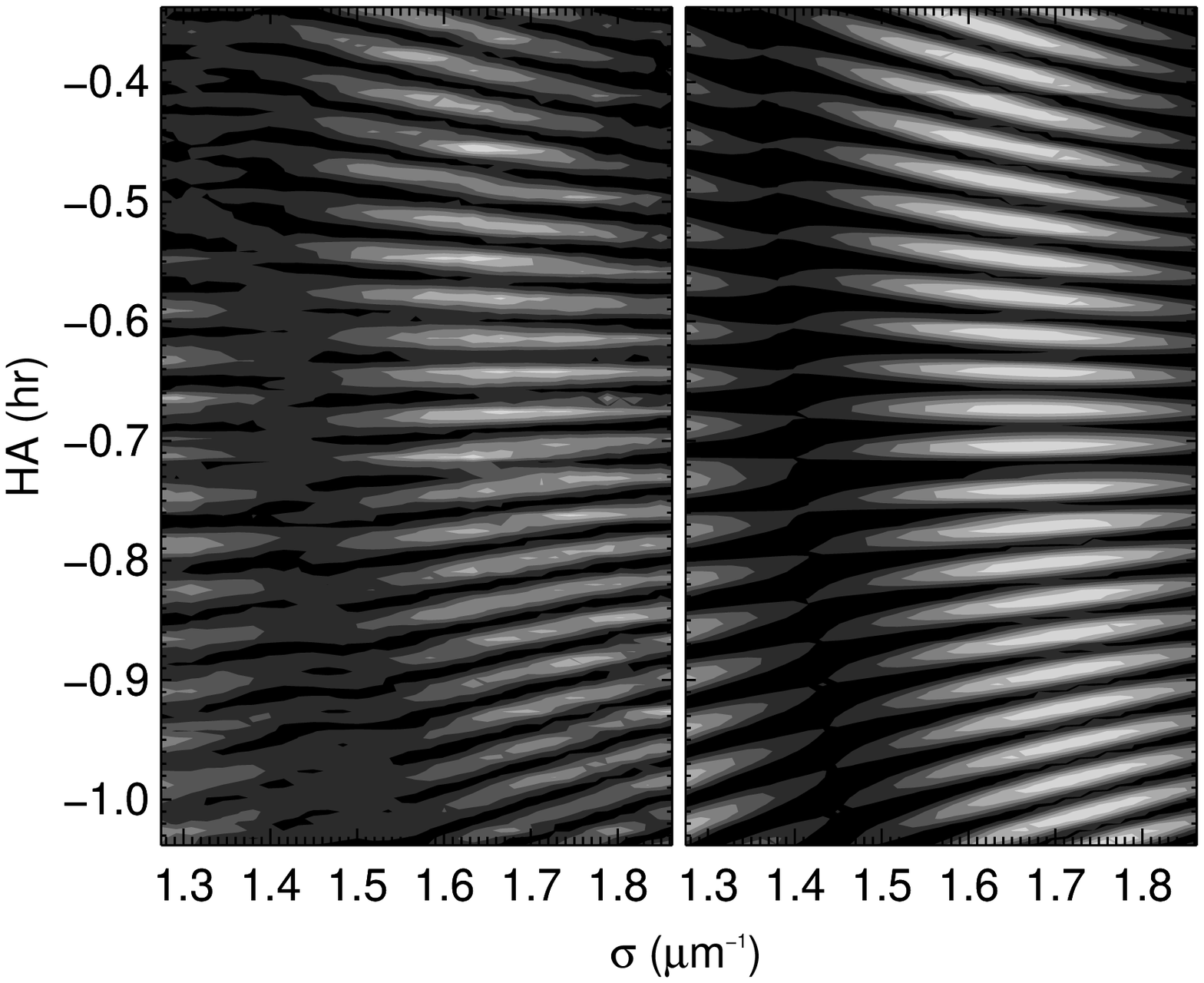}}
\subfloat[]{\includegraphics[width=0.5\textwidth]{\imgdir/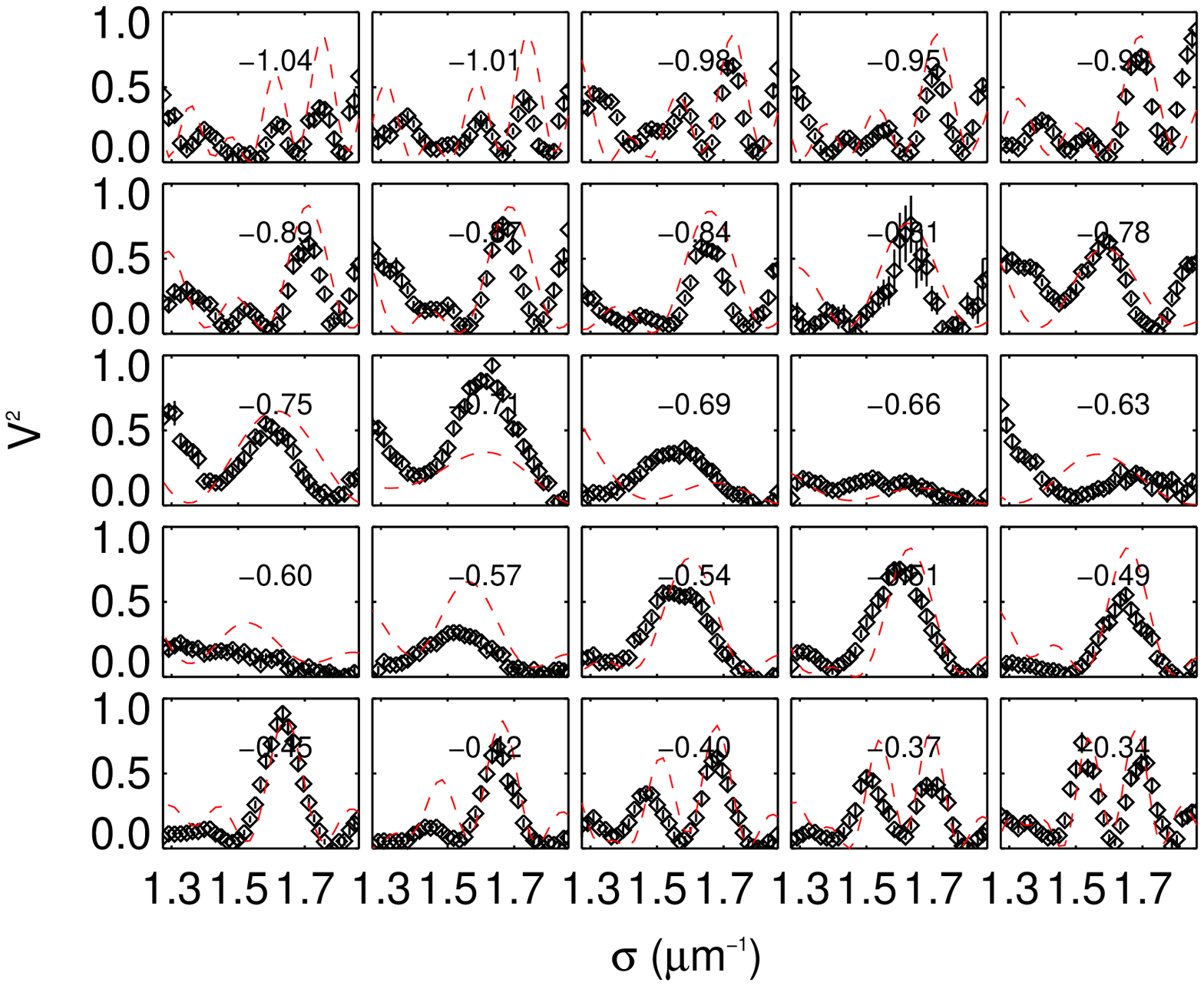}}
\caption[]
{
  Calibrated $V^2$ (left panel of (a)) of $\zeta$~Sagittarii obtained with PAVO
  on 26th July 2013 and the best fitted ternary star model (right panel of (a)).
  The grayscale in the images in (a) represents the amplitude of $V^2$. The high
  and low frequency modulations are associated with the Aa-B and Aa-Ab component
  pairs respectively. Cross-sections of (a) at different hour angles are plotted
  in (b) as $V^2$ versus wavenumber with the best fitted model represented by
  the dashed lines. Discrepancy between the best fitted model and the data as
  seen in some of the panels in (b) suggests additional astrophysical structures
  (e.g.\ different spectral type) which are not included in the triplet model.
}
\label{fig:obs_zetsgr_130726}
\end{figure}

Fig.~\ref{fig:obs_zetsgr_130726} shows the calibrated $V^2$ data obtained from
the observation and the model that best fits the data. The high frequency
modulation is contributed by the primary-secondary ($\zeta$~Sagittarii~Aa-B)
pair while the low frequency modulation (almost perpendicular to the former) is
contributed by the primary-tertiary ($\zeta$~Sagittarii~Aa-Ab) pair. The $V^2$
model of a triplet\cite{Tango:2006} used to fit the data is,
\begin{equation} \label{eq:obs_v2triplet}
\begin{split}
V_{1,2,3}^2
= \frac{1}{\left(1+\beta_2+\beta_3\right)^2}\left\{
\begin{array}{l}
V_1^2 + V_2^2\beta_2^2 + V_3^2\beta_3^2 \vspace{0.5em} \\
+ 2\sqrt{V_1^2 V_2^2}\,\beta_2\cos\phi_{1,2} + 2\sqrt{V_1^2 V_3^2}\,\beta_3\cos\phi_{1,3} \vspace{0.5em} \\
+ 2\sqrt{V_2^2 V_3^2}\,\beta_2\beta_3\cos(\phi_{1,2}-\phi_{1,3}) \\
\end{array}
\right\}
\end{split},\end{equation}
where $V_i^2$ are the visibility squared of the fringes of the individual
component stars, $\beta_i$ is the brightness ratio of the secondary or the
tertiary component with respect to the primary, and,
\begin{equation} \label{eq:obs_phi}
\begin{split}
\phi_{1,2} &= 2\pi\sigma\Delta\vec{s}_{1,2}\cdot\vec{B}, \\
\phi_{1,3} &= 2\pi\sigma\Delta\vec{s}_{1,3}\cdot\vec{B}. \\
\end{split}\end{equation}
The subscript of $\Delta\vec{s}$ represents the pair of component stars the
vector is related to. The remaining symbols in the equation were defined earlier.
The astrometric parameters extracted from the model-fitting are listed in
Table~\ref{tab:results_zetsgr}.
In the model-fitting,
all component stars were assumed to have angular diameters of $\sim$0.5mas
(uniform form disk model). The number was obtained from the discrepancy in the
total mass of the system\cite{De-Rosa:2012} and their Hipparcos
distance\cite{van-Leeuwen:2007}. Since such diameters are unresolved on a 60m
baseline, the parameters $V^2_1$, $V^2_2$ and $V^2_3$ were set to
$\sim$1 at all wavelengths.
The uncertainties of the parameters in Table~\ref{tab:results_zetsgr} are not
scaled with $\sqrt{\chi^2_R}$ and do not contain systematic errors.
The table also includes
relative astrometry from published work and ephemeris of the wider component
pair for comparison. Although the statistical uncertainties of the parameters
shown in the table are small, the true astrometric precision is limited by the
systematic errors discussed in previous section. The estimated systematic errors
in binary separation measurements are $\sim$5mas ($\sim$1\% of $\sim$0.5$''$)
for the Aa-B pair and $\sim$0.08mas ($\sim$1\% of $\sim$0.008$''$) for the Aa-Ab
pair. The estimated systematic error in position angle measurements is
$\sim$0.02$^\circ$ (the angle subtended by $\sim$2cm at a distance of 60m). 

\begin{table}
\centering
\caption{Relative astrometry of $\zeta$ Sagittarii}
\label{tab:results_zetsgr}
\begin{tabular}{l r r r r}
\toprule
Parameter & \multicolumn{1}{c}{As fitted} & \multicolumn{3}{c}{Eq=J2000} \\
\midrule
Source & This work & This work & Ephemeris$^*$ & DRS2012$\dagger$ \\
 & & & & \\
Julian yr.  & 2013.5654 & 2013.5654 & 2013.5        & 2011.31 \\
$N_{V^2}$ & 37$\times$126 & -- & -- & -- \\
LST$_X$ (Hr) & 18.396 & -- & -- & -- \\
$\beta_{\rm{Aa,B}}$  & 0.847$\pm$0.002 & -- & -- & -- \\
$\beta_{\rm{Aa,Ab}}$ & 0.778$\pm$0.002 & -- & -- & -- \\
$\Delta\alpha_{\rm{Aa,B}}$  (mas) &
-501.36$\pm$0.02 &
-501.20$\pm$0.02 &
-- & -- \\

$\Delta\delta_{\rm{Aa,B}}$  (mas) &
-42.610$\pm$0.002 &
-43.213$\pm$0.002 &
-- & -- \\

$\Delta\alpha_{\rm{Aa,Ab}}$ (mas) &
4.11$\pm$0.02 &
4.10$\pm$0.02&
-- & -- \\ 

$\Delta\delta_{\rm{Aa,Ab}}$ (mas) &
6.537$\pm$0.002	&
6.542$\pm$0.002	&
-- & -- \\

$\rho_{\rm{Aa,B}}$    (mas) &
503.17$\pm$0.02 &
503.06$\pm$0.02 &
500             &
304$\pm$4       \\

$\theta_{\rm{Aa,B}}$  ($^\circ$) &
265.142172$\pm$0.000001 &
265.072197$\pm$0.000001 &
265               &
285.59$\pm$0.17   \\

$\rho_{\rm{Aa,Ab}}$   (mas) &
7.72$\pm$0.01 &
7.72$\pm$0.01	&
-- & -- \\

$\theta_{\rm{Aa,Ab}}$ ($^\circ$) &
32.158757$\pm$0.000005 &
32.076181$\pm$0.000005 &
-- & -- \\

$\chi^2_{\rm{R}}$ & 57.8 & -- & -- & -- \\
\bottomrule
\multicolumn{5}{l}{\footnotesize{Systematic errors are not included in the
uncertainties.}} \\
\multicolumn{5}{l}{\footnotesize{$^*$based on a grade 1 orbit from Ref.~\citenum{De-Rosa:2012}}} \\
\multicolumn{5}{l}{\footnotesize{$^\dagger$Ref.~\citenum{De-Rosa:2012}}} \\
\end{tabular}
\end{table}

\section{Discussions} \label{sec:discussion}

\subsection{Astrometric error budget}

The data analysis and results demonstrate the fringe packet crossover
observation technique in performing precision astrometry on close binary stars.
A summary of systematic errors that affect the accuracy of the results is listed
in Table~\ref{tab:errbreakdown}. The errors are listed according to two
orthogonal axes in which they affect the measurements, i.e.\ along the binary
separation axis, $\delta\rho$, and along the position angle axis,
$\rho\delta\theta$. The dominant source of error in the $\rho$ axis is the
accuracy of the wavelength scale while the dominant source of error in the
$\theta$ axis is the misalignment between the imaging and the wide-angle
baselines of the interferometer (see Ref~\citenum{Woillez:2013} for
definitions). As a result, the astrometric error in the right ascension and
declination axes varies between the limits set by the two dominant error sources
as a function of $\theta$.

Errors arising from anisoplanatism of the turbulent atmosphere and time-stamping
of observation data is negligible. The former is derived with an atmospheric
coefficient\cite{Kok:2014} of $\sim$780 m$^{2/3}$s$^{1/2}$arcsec$^{-1}$ for a
total integration time of at least 10 minutes while the latter is derived from
the accuracy of the system clock of the computer used for observation. Lastly,
the contribution of the uncertainty in measuring the fringe visibility to the
astrometric error is mainly from its random error component. The astrometric
error is not sensitive to a bias in the $V^2$ measurement because the
astrometric parameters extracted from the model-fitting are dependent on the
frequency of the $V^2$ modulation. The bias, if it exists, is assumed to be
introduced by the data reduction algorithm and is wavelength and time invariant.
On the other hand, the random error of the $V^2$ measurement (usually expressed
in terms of SNR) directly affects the astrometric error through the
model-fitting process. Although it is intuitive that a lower SNR produces a
larger astrometric error, it is difficult to derive an analytical solution to
relate the two errors. Therefore, for the sake of comparison with other sources
of error, the magnitude of the astrometric error due to the random $V^2$ error
is estimated from the results in the previous section. The calibrated $V^2$
measurements have a range of average SNR between 5 to 25.


\begin{table}
\center
\caption{Breakdown of astrometric error}
\label{tab:errbreakdown}
\begin{tabular}{l c c}
\toprule
Source of errors			& $\delta\rho$	 & $\rho\delta\theta$ \\
					& ($\mu$as)	 & ($\mu$as) \\
\midrule
dOPD-related				& & \\
\midrule
Anisoplanatism				& $<$10		 & $<$10 \\
Biases in $V^2$ measurements		& -- 		 & -- \\ 
Random errors in $V^2$ measurements	& $<$400	 & $\lesssim$1 \\ 
Wavelength scale of spectrograph	& 10$\times10^3$ & -- \\
Timing accuracy				& $<$2		 & $<$2 \\
					& & \\
\midrule
Baseline-related			& & \\
\midrule
Wide-angle baseline solution		& 2		 & 2 \\
Imaging baseline alignment$^{*}$	& 200		 & 200 \\
					& \\
\midrule
Combined error 				& $\sim$10$\times10^3$ & $\sim$200 \\
\bottomrule
\multicolumn{3}{l}{\footnotesize{Estimated errors assumed:}} \\
\multicolumn{3}{l}{\footnotesize{(a) binary separation is 1$''$}} \\
\multicolumn{3}{l}{\footnotesize{(b) baseline is 100m}} \\
\multicolumn{3}{l}{\footnotesize{$^{*}$$>$20$\times$ smaller with the MUSCA
alignment}} \\
\end{tabular}
\end{table}

\subsection{Next-generation instrument}

The error budget in Table~\ref{tab:errbreakdown} shows two limiting factors for
the indirect approach method to reach higher astrometric precision. In view of
this, this subsection discusses possible improvements to the PAVO beam combiner
so that the astrometric errors determined by the two limiting factors can be
reduced. In addition, a next-generation instrument incorporating the suggested
improvements is also discussed.

Firstly, the uncertainty of the wavelength scale can be reduced by increasing
the spectral resolution of the PAVO spectrograph and by adopting a more precise
wavelength calibration procedures. With the same spectral bandwidth,
$\Delta\sigma$, an increase in spectral resolution by a factor of 100
reduces the astrometric error by a similar factor because,
\begin{equation}
  \frac{\delta\rho}{\rho}
  	\propto \frac{\delta\text{dOPD}}{\text{dOPD}}
  	\propto \frac{\delta\sigma}{\sigma}
	\propto \frac{1}{R},
\end{equation}
where $R$ denotes the resolving power of the spectrograph. Note that a high
resolution spectrum in near infrared (IR) wavelengths is less desirable because
$\Delta\sigma$ in the common $J$, $H$ and $K$ bands (0.19$\mu$m$^{-1}$,
0.11$\mu$m$^{-1}$ and 0.08$\mu$m$^{-1}$ respectively) is much smaller than in
the visible wavelengths. Even with all the near IR bands combined, the spectral
bandwidth is only half that of PAVO's. This affects the number of $V^2$
modulation cycles seen across the bandwidth for a given dOPD. A smaller number
of cycles yields a larger astrometric error from the model-fitting process. An
example of a high resolution spectro-interferometric instrument is the VEGA beam
combiner at CHARA which has a spectral resolving power of 30000 and operates in
the visible wavelengths\cite{Mourard:2009}.

The reduction in wavelength uncertainty with this approach comes at a cost of
reduced flux per spectral channel or more critically an increase in the fringe
visibility uncertainty. In the photon noise limited regime,
\begin{equation}
  \frac{\delta V^2}{V^2} = \frac{1}{\text{SNR}}
  	\propto \sqrt{R}
	\propto \frac{1}{D_{\text{tel}}\sqrt{T_{\text{exp}}}},
\end{equation}
while in the detector noise regime (when observing faint targets), $\delta
V^2/V^2 \propto R$. Such a cost can be
partially offset by the exceptional seeing condition in Antarctica, where this
observation technique is most efficient. With an atmospheric coherence time
of more than 8$\times$ that at SUSI (see Table~\ref{tab:atmcond}), the exposure
time, $T_{\text{exp}}$, of one frame of interferogram can be $\sim$10$\times$ as
long in Antarctica. The additional factor of $\sqrt{10}$ can be offset by using
siderostats with a larger diameter. Therefore, the next generation instrument
should be used with siderostats of about 60cm in diameter, $D_{\text{tel}}$, a
size similar to those used at the Palomar Testbed Interferometer (PTI) and the
Navy Precision Optical Interferometer (NPOI).

Another advantage of having a high-resolution spectrograph is the long coherence
length of the dispersed fringes. Since the standard deviation of fringe motion
due to atmospheric turbulence is expected to be $\lesssim$50$\lambda$ in
Antarctica, the dispersed fringes will always ($\sim99.7\%$ of the time) be
visible to the beam combining instrument even at a low tracking rate
($\sim$20Hz) if the spectral resolution of the spectrograph is larger than 300.

Secondly, the uncertainty of the baseline alignment can be reduced by adopting a
more precise optical alignment procedure. A procedure developed for a dedicated
astrometric beam combiner, MUSCA, at SUSI\cite{Kok:2013b} can align the
pivot point of two siderostat pupils at the beam combiner to an uncertainty of
$\sim$1mm. The pivot point of a siderostat (or a telescope) is defined as the
intersection point between two of its axes of motion, i.e.\ altitude and
azimuth. In the case of a siderostat, this point coincides with the center of
the circular aperture of the mirror. The separation of the pivot points of a
pair of siderostats (or telescopes) fundamentally defines the wide-angle
baseline of an interferometer. The pivot point of a good siderostat can be kept
from deviating from the center of the circular aperture to less than 100$\mu$m
within a time-scale of the order of a few weeks\cite{Colavita:1999,
Davis:1999a}. The amount of deviation can be characterized and is included as
part of the baseline uncertainty for the astrometric measurement. In addition,
an IR LED can be inserted into a hole centered on a siderostat, a design
similar to an astrometric siderostat of NPOI\cite{Armstrong:1998}, for real
time monitoring of the deviation if necessary.

The anisoplanatism of the turbulent atmosphere would be come the limiting factor
for higher precision astrometry once the wavelength scale calibration and the
imaging baseline alignment issues are addressed.
Table~\ref{tab:atmcond} shows the comparison of atmospheric conditions between 3
different astronomical sites.
The atmosphere in Antarctica is estimated to have $\sim$3$\times$ larger
turbulent cells (indicated by the Fried parameter, $r_0$) than the atmosphere
above SUSI.
Having such a stable atmosphere, the systematic astrometric error due to
anisoplanatism (which is inversely proportional to $r_0^{5/6}$) is negligible in
Antarctica.
Consequently, an hour long observation at SUSI would require $\sim$5 minutes (a
factor of $\sim$3$^2$ shorter in time) with the same instrument to produce the
same level of astrometric accuracy in Antarctica.

\begin{table}
\center
\caption{Typical/median atmospheric conditions$^{\dagger}$ at various long baseline interferometers}
\label{tab:atmcond}
\begin{tabular}{@{\hspace{0.6cm}}l @{\hspace{0.6cm}}c@{\hspace{0.6cm}} @{\hspace{0.6cm}}c@{\hspace{0.6cm}} @{\hspace{0.6cm}}c@{\hspace{0.6cm}}}
\toprule
Parameter		& SUSI		& VLTI		& Dome C$^{\#,*}$ \\
\midrule
Seeing ($''$)		& 1.3		& 0.86		& 0.23		\\
$r_0$ (cm)		& 7.1		& 11		& 26		\\
$\tau_0$ (ms$^{-1}$)	& 1		& 3.9		& 8.6		\\
$L_0$ (m)		& $<$80		& 22		& $\sim$8	\\
References
 & Ref.~\citenum{ten-Brummelaar:1994a,Davis:1995,Davis:1996}
 & Ref.~\citenum{Sarazin:2002,Martin:2000}
 & Ref.~\citenum{Agabi:2006,Ziad:2008} \\
\bottomrule
\multicolumn{4}{l}{\footnotesize{$^{\dagger}$specified at 500nm wavelength}} \\
\multicolumn{4}{l}{\footnotesize{$^{\#}$site used as a proxy for an interferometer in Antarctica}} \\
\multicolumn{4}{l}{\footnotesize{$^*$assumed siderostats/telescopes
are elevated ($\gtrsim$30m above ground)}} \\
\end{tabular}
\end{table}

Table~\ref{tab:nextgen} lists the features to be implemented for the next
generation instrument, which include the several upgrades discussed above. Two
additional desirable features are an extra baseline and a wider field of
view. An extra baseline, $B_2$, independent and orthogonal to the first, $B_1$,
can reduce the astrometric uncertainty by a factor of $\sqrt{1+(B_2/B_1)^{4/3}}$
in the direction of the binary separation when used simultaneously during the
fringe crossover event observation. However, the main motivation to include an
additional baseline, especially a shorter one,
is to make observable those stars which would have
A wider field
of view increases the number of observable targets. Although it does not
increase the number significantly, it would allow $\alpha$~Cen~A-B, the closest
star system to Earth and suspected to harbor a near Earth-size planet
\cite{Dumusque:2012}, to be observed. The binary separation will remain within
5$''$ for the next 5 years\cite{Pourbaix:2002}. The next epoch when the
binary separation is expected to be $<$5$''$ again is during the periastron of
the secondary component in 2035. Follow-up astrometric observations of the
binary would give a constraint on the mass of the suspected planet and the
presence of other planets further away from the host star. The additional
shorter baseline would be necessary for such observations.

\begin{table}
\center
\caption{Proposed features of an Antarctic instrument for narrow-angle astrometry}
\label{tab:nextgen}
\begin{tabular}{l c}
\toprule
Parameter			& Details/specifications \\
\midrule
Facility related		& \\
\midrule
\multirow{2}{*}{Baselines}	& 2, independent and orthogonal \\
				& 1$\times$ 10m, 1$\times$ 100m \\
Baseline alignment		& Similar to the scheme for MUSCA at SUSI \\
Light collecting elements	& 4$\times$ 60cm-diameter siderostats \\
Field of view			& $<$5$''$ \\ 
\midrule
Beam combiner related		& \\
\midrule
Optical design				& Similar to PAVO \\
Spectral bandwidth, $\Delta\sigma$	& 0.5--0.8$\mu$m (V band$^*$) \\
Spectral resolution, $R$		& $\sim$5000 (100$\times$ PAVO) \\
Number of channels, $N_{\text{ch}}$	& $\sim$2100 (100$\times$ PAVO) \\
Coherence length per channel, $L_{\text{coh}}$
					& $\sim$3mm \\
\bottomrule
\multicolumn{2}{l}{\footnotesize{$^*$see Sec.~\ref{sec:discussion} for discussion on using NIR bands}} \\
\end{tabular}
\end{table}

\section{Conclusions}

Based on the results from prototype observations with SUSI and the proposed
enhancements for a next generation instrument, the indirect approach to
narrow-angle astrometry method is capable of achieving accuracy in the regime of
tens of micro-arcseconds. This method does not require an additional metrology
system for path length measurement and uses a straightforward beam combining
instrument with no moving mechanical parts. Only the main delay line and the
siderostats, which are the essential parts of an optical long baseline
interferometer, have moving parts and therefore may require regular maintenance.
This makes the method very suitable to be deployed at a remote and poorly
accessible observation site like Antarctica.

\acknowledgments

This research was supported by the Australian Research Council's Discovery
Project funding scheme. Y.\ Kok was supported by the University of Sydney
International Scholarship (USydIS). The authors would also like to acknowledge
the use of the electronic bibliography system maintained by NASA/ADS, the
Washington Double Star Catalog maintained by the U.S. Naval Observatory and the
SIMBAD/VizieR database maintained by CDS, Strasbourg, France.

\bibliographystyle{spiebib}

\begin{thebibliography}{10}

\bibitem{Shao:1992}
Shao, M. and Colavita, M., ``{Potential of long-baseline infrared
  interferometry for narrow-angle astrometry},'' {\em \aap}~{\bf 262},
  353--358 (1992).

\bibitem{Muterspaugh:2010a}
{Muterspaugh}, M.~W., {Lane}, B.~F., {Kulkarni}, S.~R., {Konacki}, M., {Burke},
  B.~F., {Colavita}, M.~M., {Shao}, M., {Wiktorowicz}, S.~J., and {O'Connell},
  J., ``{The Phases Differential Astrometry Data Archive. I. Measurements and
  Description},'' {\em \aj}~{\bf 140},  1579--1622 (Dec. 2010).

\bibitem{Delplancke:2008}
{Delplancke}, F., ``{The PRIMA facility phase-referenced imaging and
  micro-arcsecond astrometry},'' {\em New Astronomy Review}~{\bf 52},  199--207
  (June 2008).

\bibitem{Bartko:2009}
{Bartko}, H., {Perrin}, G., {Brandner}, W., {Straubmeier}, C., {Richichi}, A.,
  {Gillessen}, S., {Paumard}, T., {Hippler}, S., {Eckart}, A., {Sch{\"o}ller},
  M., {Eisenhauer}, F., {Haubois}, X., {Lenzen}, R., {Rabien}, S.,
  {Cl{\'e}net}, Y., {Ramos}, J.~R., {Thiel}, M., {Berger}, J.~P., {Baumeister},
  H., {Kellner}, S., {Cassaing}, F., {B{\"o}hm}, A., {Hofmann}, R., {Gendron},
  E., {Klein}, R., {Dodds-Eden}, K., {Houairi}, K., {Hormuth}, F.,
  {Gr{\"a}ter}, A., {Kervella}, P., {Naranjo}, V., {Genzel}, R., {F{\'e}dou},
  P., {Henning}, T., {Hamaus}, N., {Jocou}, L., {Neumann}, U., {Haug}, M.,
  {Lacour}, S., {Laun}, W., {Kolmeder}, J., {Malbet}, F., {Rohloff}, R.,
  {Pfuhl}, O., {Perraut}, K., {Ziegleder}, J., {Rouan}, D., {Rousset}, G.,
  {Amorim}, A., and {Lima}, J., ``{GRAVITY: Astrometry on the galactic center
  and beyond},'' {\em \nar}~{\bf 53},  301--306 (Nov. 2009).

\bibitem{Pott:2009}
{Pott}, J., {Woillez}, J., {Akeson}, R.~L., {Berkey}, B., {Colavita}, M.~M.,
  {Cooper}, A., {Eisner}, J.~A., {Ghez}, A.~M., {Graham}, J.~R., {Hillenbrand},
  L., {Hrynewych}, M., {Medeiros}, D., {Millan-Gabet}, R., {Monnier}, J.,
  {Morrison}, D., {Panteleeva}, T., {Quataert}, E., {Randolph}, B., {Smith},
  B., {Summers}, K., {Tsubota}, K., {Tyau}, C., {Weinberg}, N., {Wetherell},
  E., and {Wizinowich}, P.~L., ``{Astrometry with the Keck Interferometer: The
  ASTRA project and its science},'' {\em \nar}~{\bf 53},  363--372 (Nov. 2009).

\bibitem{Kok:2013b}
{Kok}, Y., {Ireland}, M.~J., {Tuthill}, P.~G., {Robertson}, J.~G.,
  {Warrington}, B.~A., {Rizzuto}, A.~C., and {Tango}, W.~J.,
  ``{Phase-Referenced Interferometry and Narrow-Angle Astrometry with SUSI},''
  {\em \jai}~{\bf 2},  40011 (Dec. 2013).

\bibitem{Schuhler:2006}
{Schuhler}, N., {Salvad{\'e}}, Y., {L{\'e}v{\^e}que}, S., {D{\"a}ndliker}, R.,
  and {Holzwarth}, R., ``{Frequency-comb-referenced two-wavelength source for
  absolute distance measurement},'' {\em Optics Letters}~{\bf 31},  3101--3103
  (Nov. 2006).

\bibitem{Gillessen:2012}
{Gillessen}, S., {Lippa}, M., {Eisenhauer}, F., {Pfuhl}, O., {Haug}, M.,
  {Kellner}, S., {Ott}, T., {Wieprecht}, E., {Sturm}, E., {Hau{\ss}mann}, F.,
  {Kister}, C.~F., {Moch}, D., and {Thiel}, M., ``{GRAVITY: metrology},'' in
  [{\em \procspie}{\nolinebreak\hspace{0.1em}]},   {\bf 8445} (July 2012).

\bibitem{Kok:2013}
{Kok}, Y., {Ireland}, M.~J., {Robertson}, J.~G., {Tuthill}, P.~G.,
  {Warrington}, B.~A., and {Tango}, W.~J., ``{Low-cost scheme for
  high-precision dual-wavelength laser metrology},'' {\em \ao}~{\bf 52},
  2808--2814 (April 2013).

\bibitem{Tuthill:2012}
{Tuthill}, P.~G., ``Optical interferometry from the antarctic,'' in [{\em Proc.
  IAU Symp. Astrophysics from Antarctica}{\nolinebreak\hspace{0.1em}]},  (288)
  (2012).

\bibitem{Tango:2006}
{Tango}, W.~J., ``{The determination of the orbital elements of binary stars
  (Internal SUSI report)},'' {The University of Sydney} (July 2006).

\bibitem{Armstrong:2004}
{Armstrong}, J.~T., {Clark}, III, J.~H., {Gilbreath}, G.~C., {Hindsley}, R.~B.,
  {Hutter}, D.~J., {Mozurkewich}, D., and {Pauls}, T.~A., ``{Precision
  narrow-angle astrometry of binary stars with the Navy Prototype Optical
  Interferometer},'' in [{\em \procspie}{\nolinebreak\hspace{0.1em}]},   {\bf
  5491},  1700--1706 (Oct. 2004).

\bibitem{Davis:2005}
{Davis}, J., {Mendez}, A., {Seneta}, E.~B., {Tango}, W.~J., {Booth}, A.~J.,
  {O'Byrne}, J.~W., {Thorvaldson}, E.~D., {Ausseloos}, M., {Aerts}, C., and
  {Uytterhoeven}, K., ``{Orbital parameters, masses and distance to {$\beta$}
  Centauri determined with the Sydney University Stellar Interferometer and
  high-resolution spectroscopy},'' {\em \mnras}~{\bf 356},  1362--1370 (Feb.
  2005).

\bibitem{Tango:2009}
{Tango}, W.~J., {Davis}, J., {Jacob}, A.~P., {Mendez}, A., {North}, J.~R.,
  {O'Byrne}, J.~W., {Seneta}, E.~B., and {Tuthill}, P.~G., ``{A new
  determination of the orbit and masses of the Be binary system {$\delta$}
  Scorpii},'' {\em \mnras}~{\bf 396},  842--848 (June 2009).

\bibitem{Monnier:2004}
{Monnier}, J.~D., {Berger}, J.-P., {Millan-Gabet}, R., and {ten Brummelaar},
  T.~A., ``{The Michigan Infrared Combiner (MIRC): IR imaging with the CHARA
  Array},'' in [{\em New Frontiers in Stellar
  Interferometry}{\nolinebreak\hspace{0.1em}]},  {Traub}, W.~A., ed., {\em
  Society of Photo-Optical Instrumentation Engineers (SPIE) Conference Series}
  {\bf 5491},  1370 (Oct. 2004).

\bibitem{Petrov:2007}
{Petrov}, R.~G., {Malbet}, F., {Weigelt}, G., {Antonelli}, P., {Beckmann}, U.,
  {Bresson}, Y., {Chelli}, A., {Dugu{\'e}}, M., {Duvert}, G., {Gennari}, S.,
  {Gl{\"u}ck}, L., {Kern}, P., {Lagarde}, S., {Le Coarer}, E., {Lisi}, F.,
  {Millour}, F., {Perraut}, K., {Puget}, P., {Rantakyr{\"o}}, F.,
  {Robbe-Dubois}, S., {Roussel}, A., {Salinari}, P., {Tatulli}, E., {Zins}, G.,
  {Accardo}, M., {Acke}, B., {Agabi}, K., {Altariba}, E., {Arezki}, B.,
  {Aristidi}, E., {Baffa}, C., {Behrend}, J., {Bl{\"o}cker}, T., {Bonhomme},
  S., {Busoni}, S., {Cassaing}, F., {Clausse}, J.-M., {Colin}, J., {Connot},
  C., {Delboulb{\'e}}, A., {Domiciano de Souza}, A., {Driebe}, T., {Feautrier},
  P., {Ferruzzi}, D., {Forveille}, T., {Fossat}, E., {Foy}, R., {Fraix-Burnet},
  D., {Gallardo}, A., {Giani}, E., {Gil}, C., {Glentzlin}, A., {Heiden}, M.,
  {Heininger}, M., {Hernandez Utrera}, O., {Hofmann}, K.-H., {Kamm}, D.,
  {Kiekebusch}, M., {Kraus}, S., {Le Contel}, D., {Le Contel}, J.-M.,
  {Lesourd}, T., {Lopez}, B., {Lopez}, M., {Magnard}, Y., {Marconi}, A.,
  {Mars}, G., {Martinot-Lagarde}, G., {Mathias}, P., {M{\`e}ge}, P., {Monin},
  J.-L., {Mouillet}, D., {Mourard}, D., {Nussbaum}, E., {Ohnaka}, K.,
  {Pacheco}, J., {Perrier}, C., {Rabbia}, Y., {Rebattu}, S., {Reynaud}, F.,
  {Richichi}, A., {Robini}, A., {Sacchettini}, M., {Schertl}, D.,
  {Sch{\"o}ller}, M., {Solscheid}, W., {Spang}, A., {Stee}, P., {Stefanini},
  P., {Tallon}, M., {Tallon-Bosc}, I., {Tasso}, D., {Testi}, L., {Vakili}, F.,
  {von der L{\"u}he}, O., {Valtier}, J.-C., {Vannier}, M., and {Ventura}, N.,
  ``{AMBER, the near-infrared spectro-interferometric three-telescope VLTI
  instrument},'' {\em \aap}~{\bf 464},  1--12 (Mar. 2007).

\bibitem{Ireland:2008}
{Ireland}, M.~J., {M{\'e}rand}, A., {ten Brummelaar}, T.~A., {Tuthill}, P.~G.,
  {Schaefer}, G.~H., {Turner}, N.~H., {Sturmann}, J., {Sturmann}, L., and
  {McAlister}, H.~A., ``{Sensitive visible interferometry with PAVO},'' in
  [{\em \procspie}{\nolinebreak\hspace{0.1em}]},   {\bf 7013} (July 2008).

\bibitem{Robertson:2010}
{Robertson}, J.~G., {Ireland}, M.~J., {Tango}, W.~J., {Davis}, J., {Tuthill},
  P.~G., {Jacob}, A.~P., {Kok}, Y., and {Ten Brummelaar}, T.~A.,
  ``{Instrumental developments for the Sydney University Stellar
  Interferometer},'' in [{\em \procspie}{\nolinebreak\hspace{0.1em}]},   {\bf
  7734} (July 2010).

\bibitem{Levato:1987}
{Levato}, H., {Malaroda}, S., {Morrell}, N., and {Solivella}, G., ``{Stellar
  multiplicity in the Scorpius-Centaurus association},'' {\em \apjs}~{\bf 64},
  487--503 (June 1987).

\bibitem{Docobo:2006}
{Docobo}, J.~A. and {Andrade}, M., ``{A Methodology for the Description of
  Multiple Stellar Systems with Spectroscopic Subcomponents},'' {\em \apj}~{\bf
  652},  681--695 (Nov. 2006).

\bibitem{Maestro:2012}
{Maestro}, V., {Kok}, Y., {Huber}, D., {Ireland}, M.~J., {Tuthill}, P.~G.,
  {White}, T., {Schaefer}, G., {ten Brummelaar}, T.~A., {McAlister}, H.~A.,
  {Turner}, N., {Farrington}, C.~D., and {Goldfinger}, P.~J., ``{Imaging rapid
  rotators with the PAVO beam combiner at CHARA},'' in [{\em
  \procspie}{\nolinebreak\hspace{0.1em}]},   {\bf 8445} (July 2012).

\bibitem{Tokovinin:2010}
{Tokovinin}, A., {Mason}, B.~D., and {Hartkopf}, W.~I., ``{Speckle
  Interferometry at the Blanco and SOAR Telescopes in 2008 and 2009},'' {\em
  \aj}~{\bf 139},  743--756 (Feb. 2010).

\bibitem{Hartkopf:2012}
{Hartkopf}, W.~I., {Tokovinin}, A., and {Mason}, B.~D., ``{Speckle
  Interferometry at SOAR in 2010 and 2011: Measures, Orbits, and Rectilinear
  Fits},'' {\em \aj}~{\bf 143},  42 (Feb. 2012).

\bibitem{Tokovinin:2012}
{Tokovinin}, A., ``{Speckle Interferometry and Orbits of ``Fast'' Visual
  Binaries},'' {\em \aj}~{\bf 144},  56 (Aug. 2012).

\bibitem{Heintz:1990}
{Heintz}, W.~D., ``{Orbits of 15 visual binaries},'' {\em \aaps}~{\bf 82},
  65--69 (Jan. 1990).

\bibitem{De-Rosa:2012}
{De Rosa}, R.~J., {Patience}, J., {Vigan}, A., {Wilson}, P.~A., {Schneider},
  A., {McConnell}, N.~J., {Wiktorowicz}, S.~J., {Marois}, C., {Song}, I.,
  {Macintosh}, B., {Graham}, J.~R., {Bessell}, M.~S., {Doyon}, R., and {Lai},
  O., ``{The Volume-limited A-Star (VAST) survey - II. Orbital motion
  monitoring of A-type star multiples},'' {\em \mnras}~{\bf 422},  2765--2785
  (June 2012).

\bibitem{van-Leeuwen:2007}
{van Leeuwen}, F., ``{Validation of the new Hipparcos reduction},'' {\em
  \aap}~{\bf 474},  653--664 (Nov. 2007).

\bibitem{Woillez:2013}
{Woillez}, J. and {Lacour}, S., ``{Wide-angle, Narrow-angle, and Imaging
  Baselines of Optical Long-baseline Interferometers},'' {\em \apj}~{\bf 764},
  109 (Feb. 2013).

\bibitem{Kok:2014}
{Kok}, Y., {\em {Phase-referencing interferometry and narrow-angle astrometry
  with SUSI}}, PhD thesis, The University of Sydney (2014).

\bibitem{Mourard:2009}
{Mourard}, D., {Clausse}, J.~M., {Marcotto}, A., {Perraut}, K., {Tallon-Bosc},
  I., {B{\'e}rio}, P., {Blazit}, A., {Bonneau}, D., {Bosio}, S., {Bresson}, Y.,
  {Chesneau}, O., {Delaa}, O., {H{\'e}nault}, F., {Hughes}, Y., {Lagarde}, S.,
  {Merlin}, G., {Roussel}, A., {Spang}, A., {Stee}, P., {Tallon}, M.,
  {Antonelli}, P., {Foy}, R., {Kervella}, P., {Petrov}, R., {Thiebaut}, E.,
  {Vakili}, F., {McAlister}, H., {ten Brummelaar}, T., {Sturmann}, J.,
  {Sturmann}, L., {Turner}, N., {Farrington}, C., and {Goldfinger}, P.~J.,
  ``{VEGA: Visible spEctroGraph and polArimeter for the CHARA array: principle
  and performance},'' {\em \aap}~{\bf 508},  1073--1083 (Dec. 2009).

\bibitem{Colavita:1999}
Colavita, M., Wallace, J., Hines, B., Gursel, Y., Malbet, F., Palmer, D., Pan,
  X., Shao, M., Yu, J., Boden, A., and Others, ``{The Palomar testbed
  interferometer},'' {\em \apj}~{\bf 510},  505--521 (1999).

\bibitem{Davis:1999a}
{Davis}, J., {Tango}, W.~J., {Booth}, A.~J., {Thorvaldson}, E.~D., and
  {Giovannis}, J., ``{The Sydney University Stellar Interferometer - II.
  Commissioning observations and results},'' {\em \mnras}~{\bf 303},  783--791
  (Mar. 1999).

\bibitem{Armstrong:1998}
{Armstrong}, J.~T., {Mozurkewich}, D., {Rickard}, L.~J., {Hutter}, D.~J.,
  {Benson}, J.~A., {Bowers}, P.~F., {Elias}, II, N.~M., {Hummel}, C.~A.,
  {Johnston}, K.~J., {Buscher}, D.~F., {Clark}, III, J.~H., {Ha}, L., {Ling},
  L., {White}, N.~M., and {Simon}, R.~S., ``{The Navy Prototype Optical
  Interferometer},'' {\em \apj}~{\bf 496},  550--571 (Mar. 1998).

\bibitem{ten-Brummelaar:1994a}
{ten Brummelaar}, T., ``{Taking the Twinkle Out of the Stars: an Adaptive
  Wavefront Tilt Correction Servo and Preliminary Seeing Study for SUSI},''
  {\em \pasp}~{\bf 106},  915 (Aug. 1994).

\bibitem{Davis:1995}
{Davis}, J., {Lawson}, P.~R., {Booth}, A.~J., {Tango}, W.~J., and
  {Thorvaldson}, E.~D., ``{Atmospheric path variations for baselines up to 80m
  measured with the Sydney University Stellar Interferometer},'' {\em
  \mnras}~{\bf 273},  L53--L58 (Apr. 1995).

\bibitem{Davis:1996}
{Davis}, J. and {Tango}, W., ``{Measurement of the Atmospheric Coherence
  Time},'' {\em \pasp}~{\bf 108},  456--+ (May 1996).

\bibitem{Sarazin:2002}
{Sarazin}, M. and {Tokovinin}, A., ``{The Statistics of Isoplanatic Angle and
  Adaptive Optics Time Constant derived from DIMM Data},'' in [{\em European
  Southern Observatory Conference and Workshop
  Proceedings}{\nolinebreak\hspace{0.1em}]},  {Vernet}, E., {Ragazzoni}, R.,
  {Esposito}, S., and {Hubin}, N., eds., {\em European Southern Observatory
  Conference and Workshop Proceedings} {\bf 58},  321 (2002).

\bibitem{Martin:2000}
{Martin}, F., {Conan}, R., {Tokovinin}, A., {Ziad}, A., {Trinquet}, H.,
  {Borgnino}, J., {Agabi}, A., and {Sarazin}, M., ``{Optical parameters
  relevant for High Angular Resolution at Paranal from GSM instrument and
  surface layer contribution},'' {\em \aaps}~{\bf 144},  39--44 (May 2000).

\bibitem{Agabi:2006}
{Agabi}, A., {Aristidi}, E., {Azouit}, M., {Fossat}, E., {Martin}, F.,
  {Sadibekova}, T., {Vernin}, J., and {Ziad}, A., ``{First Whole Atmosphere
  Nighttime Seeing Measurements at Dome C, Antarctica},'' {\em \pasp}~{\bf
  118},  344--348 (Feb. 2006).

\bibitem{Ziad:2008}
{Ziad}, A., {Aristidi}, E., {Agabi}, A., {Borgnino}, J., {Martin}, F., and
  {Fossat}, E., ``{First statistics of the turbulence outer scale at Dome C},''
  {\em \aap}~{\bf 491},  917--921 (Dec. 2008).

\bibitem{Dumusque:2012}
{Dumusque}, X., {Pepe}, F., {Lovis}, C., {S{\'e}gransan}, D., {Sahlmann}, J.,
  {Benz}, W., {Bouchy}, F., {Mayor}, M., {Queloz}, D., {Santos}, N., and
  {Udry}, S., ``{An Earth-mass planet orbiting {$\alpha$} Centauri B},'' {\em
  \nat}~{\bf 491},  207--211 (Nov. 2012).

\bibitem{Pourbaix:2002}
{Pourbaix}, D., {Nidever}, D., {McCarthy}, C., {Butler}, R.~P., {Tinney},
  C.~G., {Marcy}, G.~W., {Jones}, H.~R.~A., {Penny}, A.~J., {Carter}, B.~D.,
  {Bouchy}, F., {Pepe}, F., {Hearnshaw}, J.~B., {Skuljan}, J., {Ramm}, D., and
  {Kent}, D., ``{Constraining the difference in convective blueshift between
  the components of alpha Centauri with precise radial velocities},'' {\em
  \aap}~{\bf 386},  280--285 (Apr. 2002).

\end{thebibliography}


\end{document}